# The golden age of solar magnetography at Paris-Meudon observatory in the second half of the twentieth century

*Jean-Marie Malherbe (emeritus astronomer)*

Observatoire de Paris, PSL Research University, LIRA, France - Email: Jean-Marie.Malherbe@obspm.fr
ORCID: https://orcid.org/0000-0002-4180-3729

**ABSTRACT**

This paper describes advances in solar magnetography and developments in instrumental techniques of polarimetry and spectroscopy made at Paris-Meudon observatory in the second half of the twentieth century. The adventure started from Lyot expertise and extended progressively to the measurement of vector magnetic fields using various and improving polarimetric techniques (such as beam exchange or grid) or new spectroscopic methods (such as the MSDP imaging slicer), at Meudon and Pic du Midi, ending by the achievement of the state-of-the-art optimized and polarization free telescope THEMIS in 1999.

**KEYWORDS**

Sun, magnetograph, magnetic field, photosphere, chromosphere, Stokes parameters, history

**INTRODUCTION**

Why are magnetic fields so important in solar physics ? It is known since the discovery by George Hale (1908) of the magnetic nature of sunspots that magnetic fields play a major role in solar activity. The 11-year cycle, characterized by flares and coronal mass ejections, is related to the 22-year magnetic cycle (Hale & Nicholson, 1925), with polarity reversal of the solar dipole at each activity maximum. Magnetic energy is stored in active regions (sunspots, faculae); it is released by magnetic reconnection and partly converted into kinetic energy (ejections, mass motions) and heat (brightening, UV radiation) in flares. A detailed history of solar magnetic fields since Hale was written by Stenflo (2015). Intense magnetic fields are detected by the Zeeman effect on spectral lines, while weak, turbulent and unresolved fields of the quiet Sun were more recently evidenced by the Hanle effect. It is in all cases necessary to measure the Stokes vector (I, Q, U, V) where I is the intensity and Q, U, V contain the linear and circular polarization state of light. Stenflo (1994, 2013) and Landi (1992) reviewed and discussed how magnetic fields can be revealed by polarimetric and spectroscopic measurements.

**I – THE PIONEERS AT MEUDON**

Some researchers of Meudon observatory belong to the pioneers in polarimetry, such as Bernard Lyot (1897-1952). The invention of the coronagraph made him famous in the thirties, and he soon invented the monochromatic polarizing filter (Lyot, 1944). It is composed of optical stages based on the interference of ordinary and extraordinary rays in birefringent crystals, such as quartz or calcite (figures 1, 2 & 3).

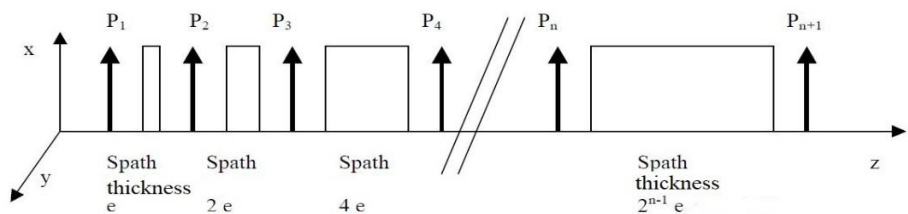

*Figure 1 : A n-stage Lyot filter made of n+1 parallel polarizers and n birefringent crystals (optical axis at 45° of the polarizers). The thickness of plate n is $2^{n-1} e$ (e = thickness of plate 1). Such a filter provides the channeled transmission of Figure 2. Courtesy OP.*

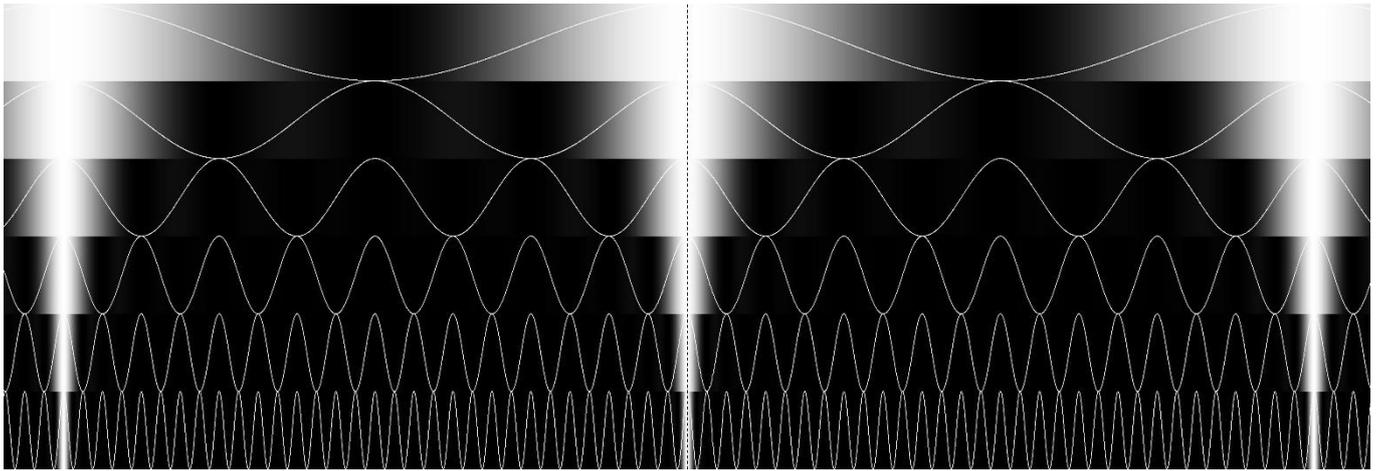

*Figure 2 : A 6 stage Lyot filter centred on Hα 6562.8 Å line, FWHM 0.25 Å (from top to bottom, the transmission of stages 1 to 6); the distance between peaks is 16 Å; the main peak (dashed line) must be isolated by a narrow interference filter (5-10 Å FWHM typical, not shown). Courtesy OP.*

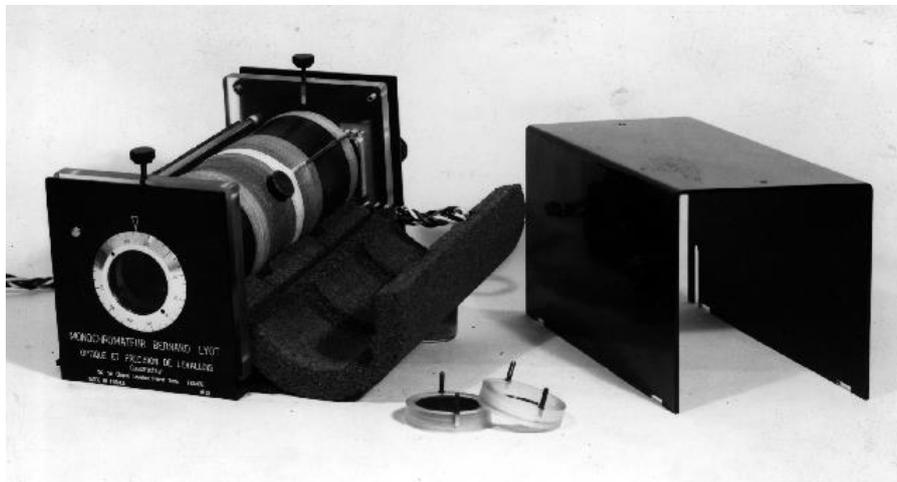

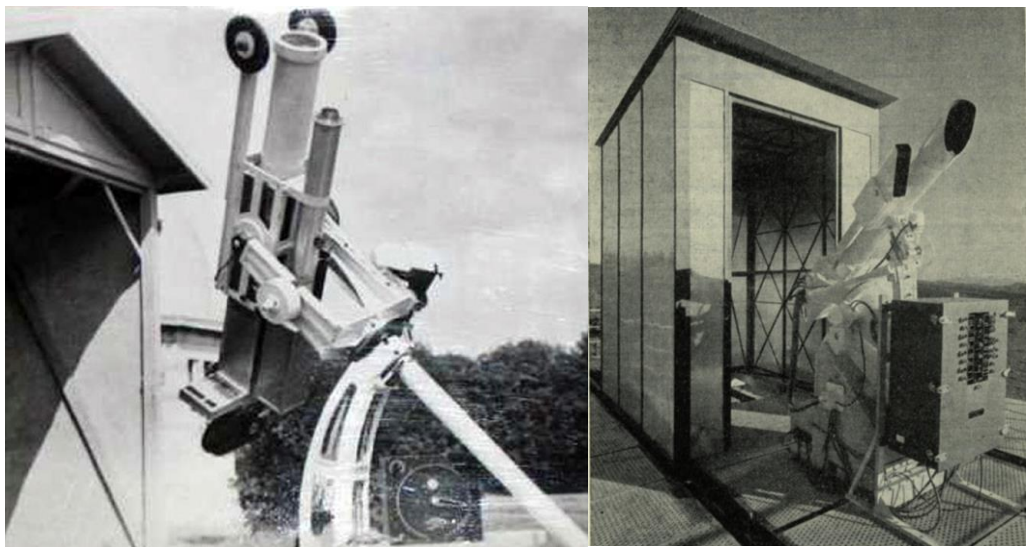

*Figure 3 : the Lyot filter (top) was mounted in 1954 on the Meudon Hα heliograph (left) and was reproduced by the french company OPL to be disseminated in many places around the world, such as Haute Provence Observatory (OHP) for the International Geophysical Year (1957-58). Courtesy OP.*

When the polarizer $P_{n+1}$ of figure 1 is rotated and directed along Oy axis instead of Ox, one obtain the result of Figure 4. The transmission peak, centred in a spectral line (here the forbidden coronal green line of FeXIV 5303 Å) is replaced by two adjacent peaks in the continuum. This is the principle of the coronameter which allowed monochromatic imagery of a coronal line and of the continuum around, the ratio of which providing the line intensity. Such a coronameter could be combined with a polarimeter and some instruments were developed, after the death of Lyot, by his collaborator Audoin Dollfus (1924-2010) and later by Pierre Charvin (1931-1990) or Jean Arnaud (1945-2010).

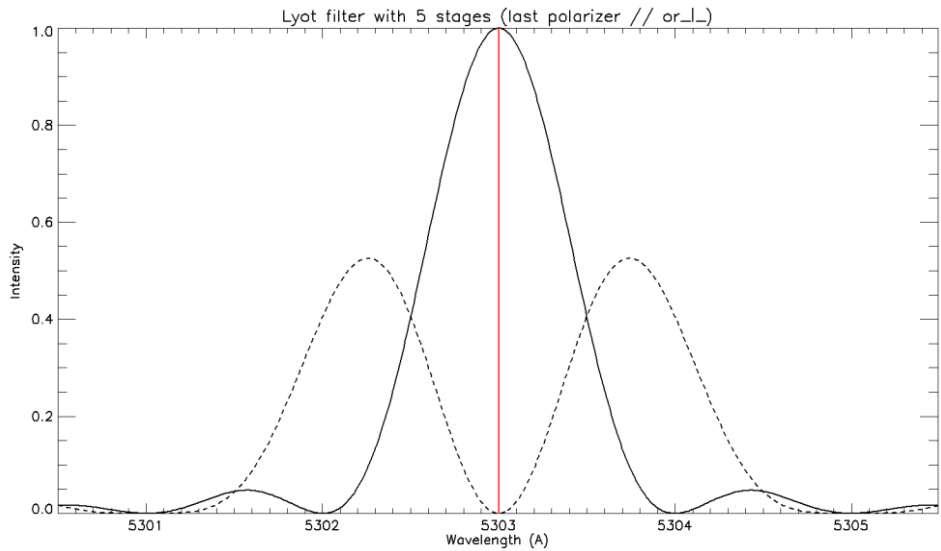

*Figure 4 : the coronameter is a Lyot filter with two crossed positions of the exit polarizer, allowing to measure alternatively the line intensity (transmission in solid line) or the adjacent continuum (transmission in dashed line). Courtesy OP.*

In order to determine the polarization of light or spectral lines, one needs a polarimeter. Let us consider a device made of a wave plate of retardance δ, with fast and slow axis rotated by α angle with respect to Ox and Oy, and a polarizer either along Ox or Oy (figure 5).

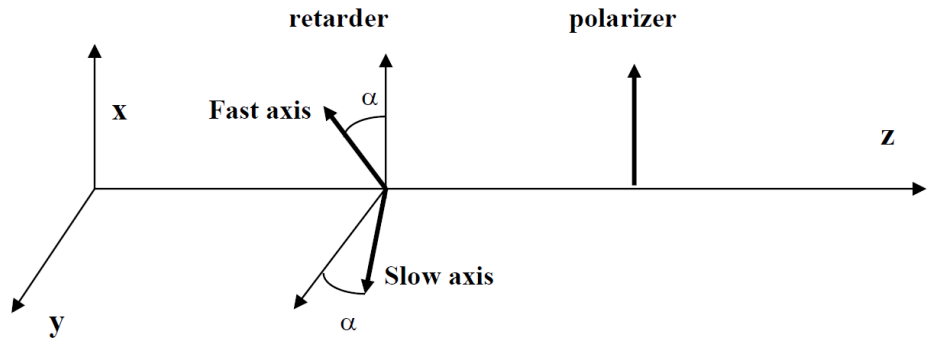

*Figure 5 : A simple polarimeter made of a rotating waveplate and a polarizer*

The Stokes vector **S**$_{out}$ is related to the Stokes vector **S**$_{in}$ by the Müller matrices product :

**S**$_{out}$ = **P R**$_{-\alpha}$ **T**$_\delta$ **R**$_\alpha$ **S**$_{in}$ where P, R, T are respectively the polarizer, rotation and retardance matrices. The output intensity is given by :

$$I_{out} = \tfrac{1}{2} [\, I_{in} \pm \{ Q_{in}(\cos^2(2\alpha) + \sin^2(2\alpha)\cos(\delta)) + U_{in}\sin(4\alpha)\sin^2(\delta/2) - V_{in}\sin(\delta)\sin(2\alpha) \} \,]$$

The two signs correspond to polarizers either along Ox or Oy, or to a polarizing beam shifter-splitter, such as a birefringent Wollaston prism. With δ = π (half waveplate), $I_{out} = \tfrac{1}{2}[\, I_{in} \pm \{Q_{in}\cos(4\alpha) + U_{in}\sin(4\alpha)\} \,]$. The sum is $I_{in}$ and the difference is $\Delta I = Q_{in}\cos(4\alpha) + U_{in}\sin(4\alpha)$. Hence, $Q_{in}$ and $U_{in}$ are modulated by $\cos(4\alpha)$ and $\sin(4\alpha)$.

For α varying by steps, for instance α = 0, π/8, π/4, 3π/8, π/2, ..., one finally gets ΔI = $Q_{in}$, $V_{in}$, -$Q_{in}$, -$V_{in}$, $Q_{in}$ ... With a half waveplate rotating at constant speed, such that α = ωt, an electronic demodulation can be undertaken by multiplying the signal ΔI either by cos(4ωt) or sin(4ωt); time integration of ΔI cos(4ωt) gives ½ $Q_{in}$T while time integration of ΔI sin(4ωt) provides ½$U_{in}$T where T is the rotation period. When a quarter waveplate is inserted in front of the half waveplate, with axis parallel to Ox and Oy, it allows to exchange U and V, so that we have now $I_{out}$ = ½ [ $I_{in}$ ± {$Q_{in}$ cos(4α) + $V_{in}$ sin(4α)} ] ; the sum is $I_{in}$ and the difference ΔI = $Q_{in}$ cos(4α) + $V_{in}$ sin(4α). With α = ωt, time integration of ΔI cos(4ωt) gives ½ $Q_{in}$T while time integration of ΔI sin(4ωt) gives ½$V_{in}$T.

The photoelectric polarimeter follows this principle and was described by Dofffus (1958); it allowed to analyse the linear polarization of the light, possibly in combination with a coronameter. The light crossed a birefringent prism which delivered two beams of orthogonally polarized light. Intensities were converted in currents by two photoelectric cells. A modulator, made of a rotating half waveplate, produced varying currents which were analyzed by a synchronous electronic demodulator in order to separate the modulated Stokes parameters Q and U of the constant intensity I. Two of these instruments were set up at Meudon and Pic du Midi in 1963. A continuous survey occurred during one solar cycle (number 20): it was coordinated by Meudon until 1967 and Pic du Midi after. Charvin (1965) studied the polarization of forbidden lines such as the FeXIV 5303 Å line and presented observations made with the polarimeter of Figure 6, which was improved later for the Pic du Midi (Charvin, 1971) and intensively used by Arnaud (1982) to detect the orientation of coronal magnetic fields.

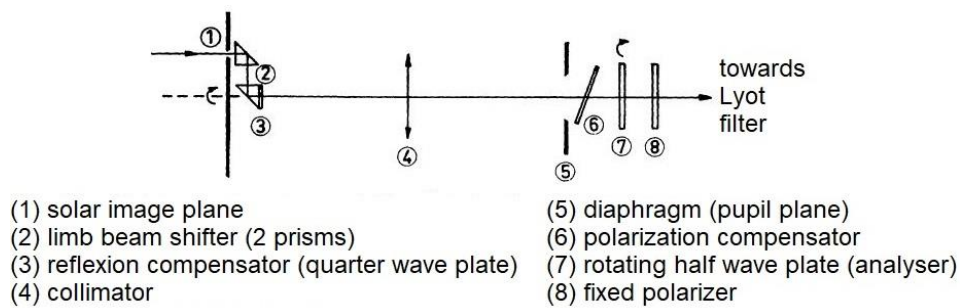

(1) solar image plane
(2) limb beam shifter (2 prisms)
(3) reflexion compensator (quarter wave plate)
(4) collimator
(5) diaphragm (pupil plane)
(6) polarization compensator
(7) rotating half wave plate (analyser)
(8) fixed polarizer

*Figure 6 : the polarimeter used by Charvin (1965) at Meudon to study the linear polarization of coronal forbidden lines; it could work with a broad band filter or with a monochromatic Lyot filter. Courtesy OP.*

Leroy (1962) used the polarimeter of Figure 7 to study the broad band linear polarization in sunspots and was able to determine the strength and direction of the transverse magnetic field in regions where the solar spectrum is filled with many lines sensitive to the Zeeman effect (such as the blue/green). He initiated at Meudon the magnetography of active regions before moving to Pic du Midi; his career was then fully devoted to solar and stellar polarimetry (Leroy, 1998 book).

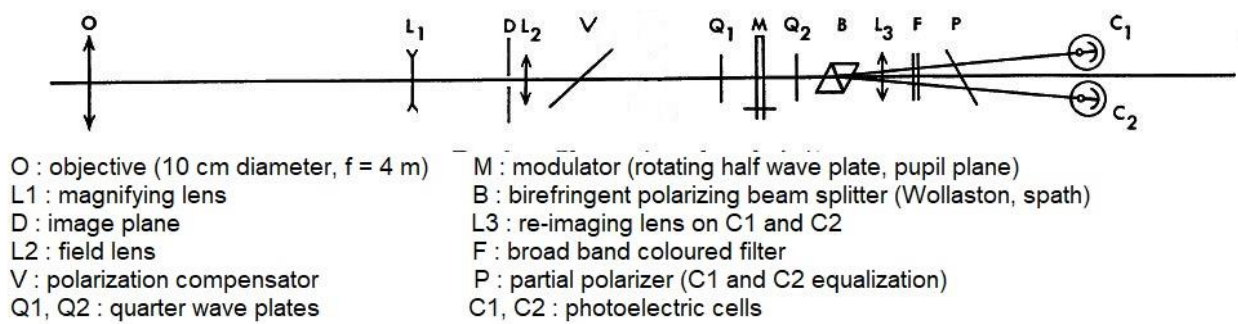

O : objective (10 cm diameter, f = 4 m)
L1 : magnifying lens
D : image plane
L2 : field lens
V : polarization compensator
Q1, Q2 : quarter wave plates
M : modulator (rotating half wave plate, pupil plane)
B : birefringent polarizing beam splitter (Wollaston, spath)
L3 : re-imaging lens on C1 and C2
F : broad band coloured filter
P : partial polarizer (C1 and C2 equalization)
C1, C2 : photoelectric cells

*Figure 7 : the polarimeter used by Leroy (1962) at Meudon to study the linear polarization of sunspots with broad band filters. The Wollaston delivered a dual beam feeding photoelectric cells. Courtesy OP.*

Dollfus (1985) built a tunable monochromatic filter (the FPSS, figure 8) which was developed and tested at Meudon and then moved in 1992 to the Pic du Midi Turret-Dome. It was a complex assembly of two Lyot-type filters and a polarimeter (0.13 Å FWHM at 5800 Å). The velocity and magnetic fields of fine structures, in spectral lines such as FeI 5576 Å (for velocities), FeI 6173 Å or Na D1 5896 Å (for magnetic fields, Figure 9), were studied in collaboration with R. Muller and J. Moity.

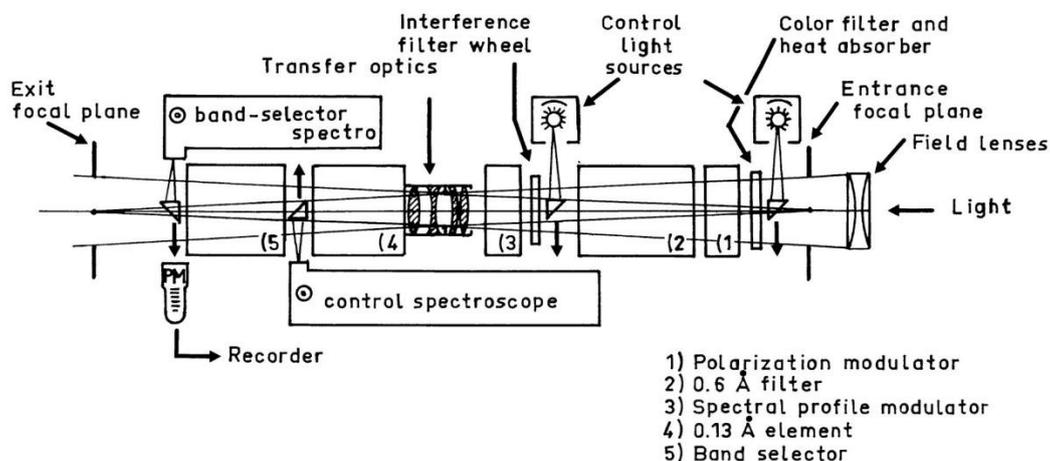

*Figure 8: The "Filtre Polarisant Solaire Selectif" (FPSS) developed at Meudon and installed on the Turret-Dome in 1992 was composed of two Lyot filters and a polarization modulator. Courtesy OP.*

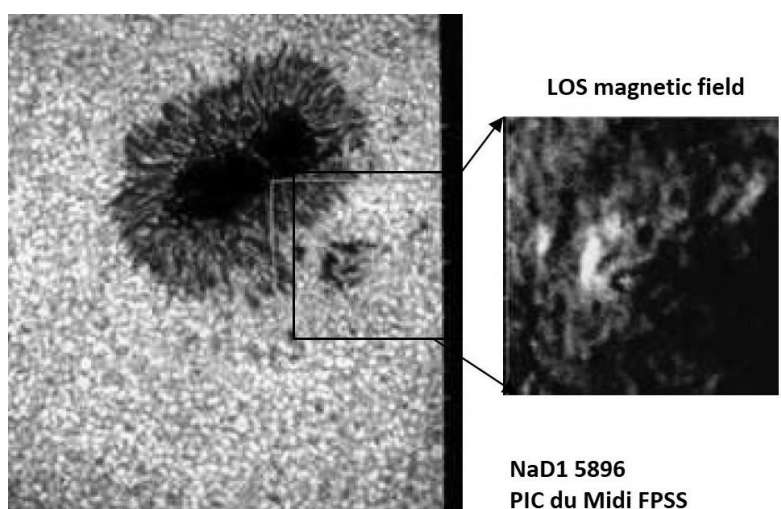

*Figure 9: Magnetography of a sunspot in NaD1 line with the FPSS developed by Dollfus. Courtesy OP.*

## II - POLARIMETRY WITH THE HALE GRID: THE FLARE SPECTROGRAPH (1961) AND THE PCM SURVEY OF MAGNETIC ACTIVITY (1962)

Systematic observations of line of sight (LOS) magnetic fields started in Meudon in October 1962 (Michard & Rayrole, 1965) ; a PCM ("Plaque de Champ Magnétique" or Magnetic Field Plate) of 9 x 24 cm² (Figure 10) of active regions was produced every sunny day using the large 7-metre research spectrograph of Lucien d'Azambuja in the Zeeman sensitive magnetic lines of FeI 6302 Å (Lande factor $g^* = 2.5$) or FeI 5250 Å ($g^* = 3.0$). The polarimeter was a simple Hale-Nicholson grid (1938) located upon the entrance slit with alternatively strips of quarter and three quarter waveplates and a polaroid. The Hale grid allowed to record spectra of I+V and I-V (Figure 10) on regions separated by 3.68" (grid step of 3.2"), so that the measurements were not co-spatial. The field of view (FOV) was 7' x 9'. The spectral bandpass was 0.52 Å. The telescope was a Newtonian system of 40 cm aperture and 7 m focal length, with a magnifying secondary mirror providing an equivalent focal length of 25 m. It was fed by the Foucault siderostat of 75 cm diameter (Figure 11). The focal length of the spectrograph was also 7 m; it used a 1200 grooves/mm grating providing the dispersion of about 3.85 mm/Å at order 2, so that 110 spectra (each 2 mm wide) were recorded on the PCM to cover the active region. The LOS magnetic field strength $B_{//}$ was deduced from the Zeeman splitting (2w) of Figure 10; w was precisely measured with a "lambdameter" and was proportional to $B_{//}$ by the formula : $w = 4.67 \cdot 10^{-13} B_{//} g^* \lambda^2$ with $B_{//}$ in Gauss and λ in Å.

The "lambdameter" had a punching paper tape recorder, that was later read by the IBM1401 computer of the observatory to print listings of coded magnetic maps such as the one of Figure 12.

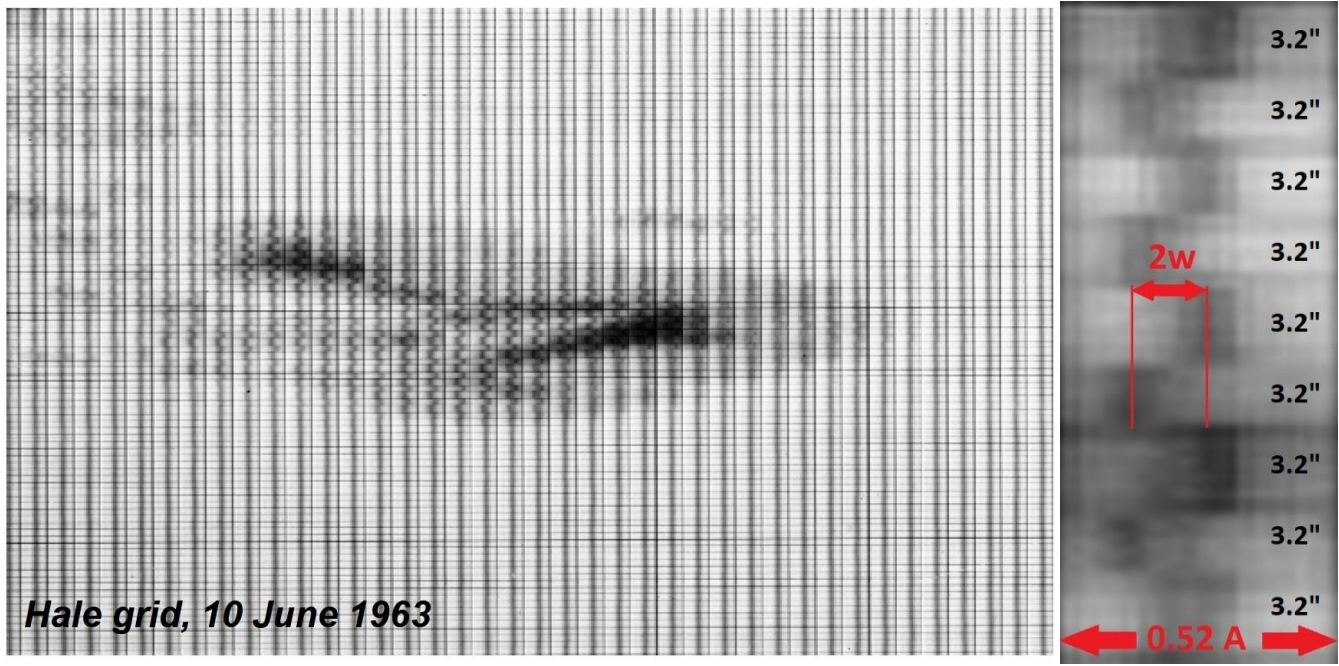

Figure 10: the magnetic field plate of 10 June 1963 with juxtaposed spectra. Courtesy OP.

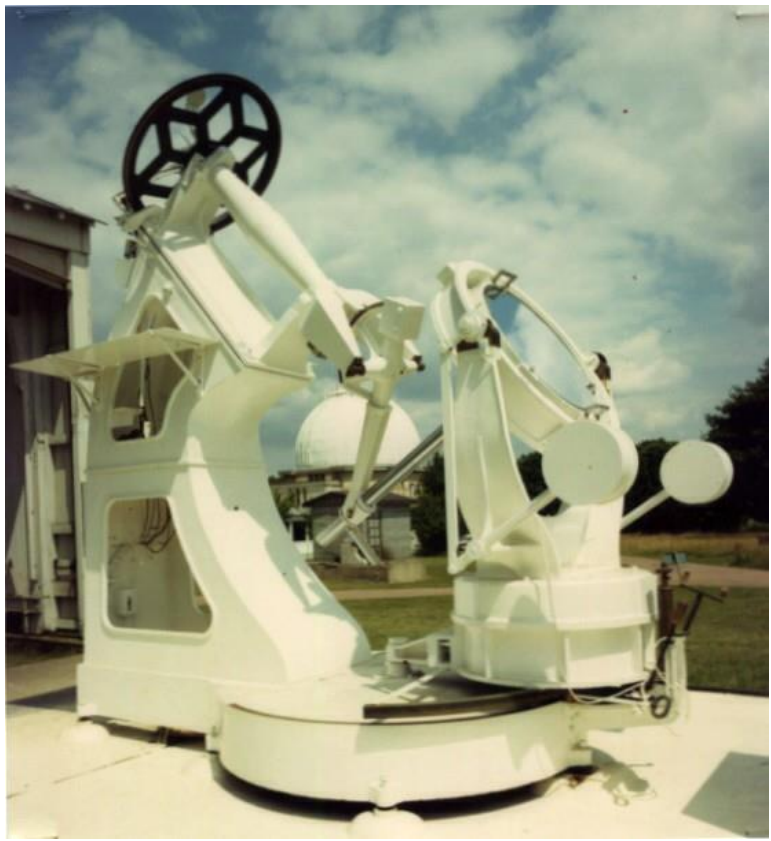

Figure 11: the 75 cm Foucault siderostat catching the Sun. Courtesy OP.

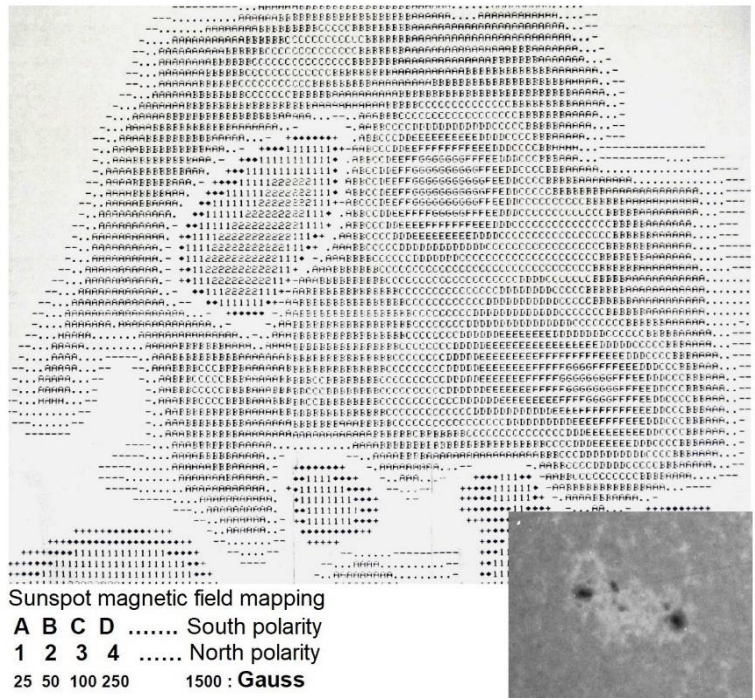

*Figure 12: coded map of magnetic fields printed by the IBM1401 computer. Courtesy OP.*

Michard *et al* (1961) observed the LOS magnetic field in an active region, before and during a flare (23 April 1960), using the Hale grid at the 9-metre spectrograph of Pic du Midi. They used the 50 cm aperture telescope (equivalent focal length of 22 m), the CaI 6103 Å ($g^* = 2.0$) line and the 600 grooves/mm grating providing the dispersion of 3.1 mm/Å. The grid step was 3.8". However, there was no permanent staff at Pic du Midi to perform such observations on a regular basis.

## III - THE SUCCESSIVE VERSIONS OF MEUDON MAGNETOGRAPH

Jean Rayrole (1932-2008) replaced the Hale grid by a more sophisticated system (figure 13) composed of a quarter waveplate $Q_2$ (quartz) and a rotating polarizer $P_1$, allowing to record on the photographic plate $P_2$ co-spatial spectra of I+V and I-V; however, these spectra were note simultaneous, because the polarizer had to rotate between two consecutive exposures. $Q_2$ was temperature controlled to stay precisely quarter wave either for the wavelengths of FeI 6302 Å or FeI 5250 Å. $Q_1$ was a half waveplate (polarization rotator) to correct the parasitic polarization induced by the reflexion on the siderostat mirror. The surface of the active region was scanned by the telescope $M_2$ and the spectra were formed on the emulsion $P_2$ which was translating in perfect synchronism with $M_2$. Rayole (thesis, 1967) intensively studied the magnetic structure of sunspots with this instrument.

The magnetograph was modernized in order to observe simultaneously the co-spatial spectra I+V and I-V. For that purpose, the rotating polarizer $P_1$ of Figure 13 was replaced by a birefringent crystal of calcite. It was possible to observe simultaneously three lines, the magnetic FeI 5225 Å ($g^* = 2.25$) and 5250 Å ($g^* = 3$), and the non magnetic FeI 5576 Å ($g^* = 0$). At the end of the seventies, the photographic plate was replaced by Reticon linear arrays with a Texas Instrument micro-computer, and the instrument was exploited by Jacques Moity in the frame of coordinated campaigns with other instruments such as the Solar Maximum Mission satellite (NASA) launched in 1980, or other ground based instruments. A colour image processor, a Comtal Vision one/20, was bought in 1979 to display the magnetic field maps (Figure 14) calculated by the Texas computer and sharing data with it owing to the 8-inch/512 Ko floppy disks. Meanwhile, Rayrole was busy by the design of a new, powerful and optimized vector magnetograph, without any source of instrumental polarization, the future THEMIS telescope in Tenerife which was commissioned in 1999.

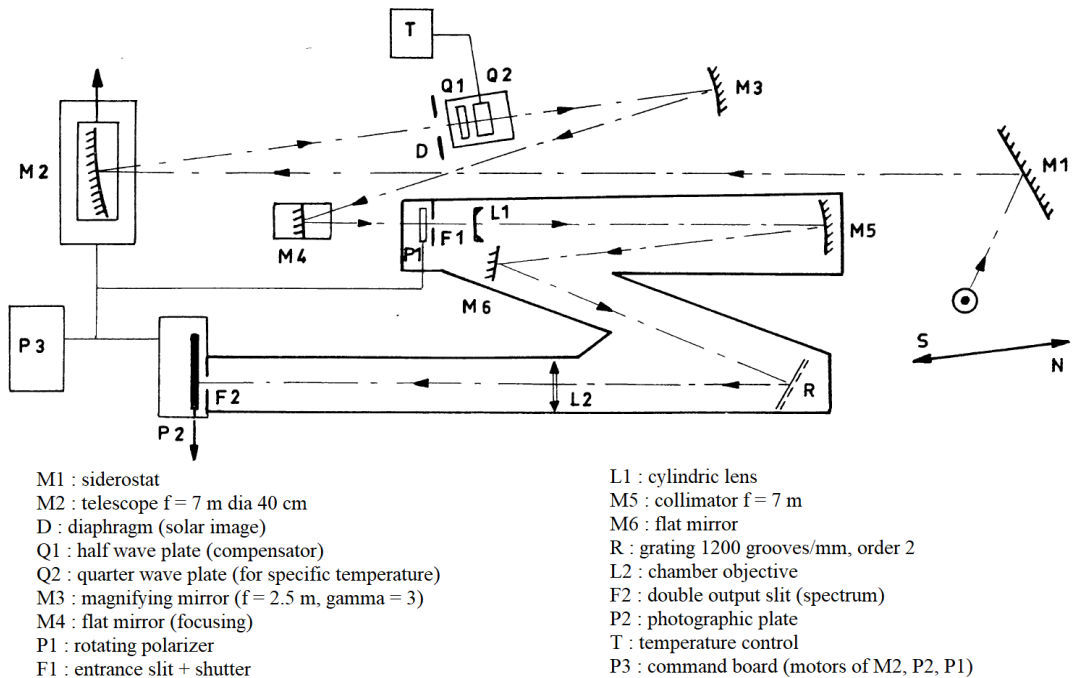

M1 : siderostat
M2 : telescope f = 7 m dia 40 cm
D : diaphragm (solar image)
Q1 : half wave plate (compensator)
Q2 : quarter wave plate (for specific temperature)
M3 : magnifying mirror (f = 2.5 m, gamma = 3)
M4 : flat mirror (focusing)
P1 : rotating polarizer
F1 : entrance slit + shutter

L1 : cylindric lens
M5 : collimator f = 7 m
M6 : flat mirror
R : grating 1200 grooves/mm, order 2
L2 : chamber objective
F2 : double output slit (spectrum)
P2 : photographic plate
T : temperature control
P3 : command board (motors of M2, P2, P1)

*Figure 13 : the Meudon magnetograph. After Rayrole (1967). Courtesy OP.*

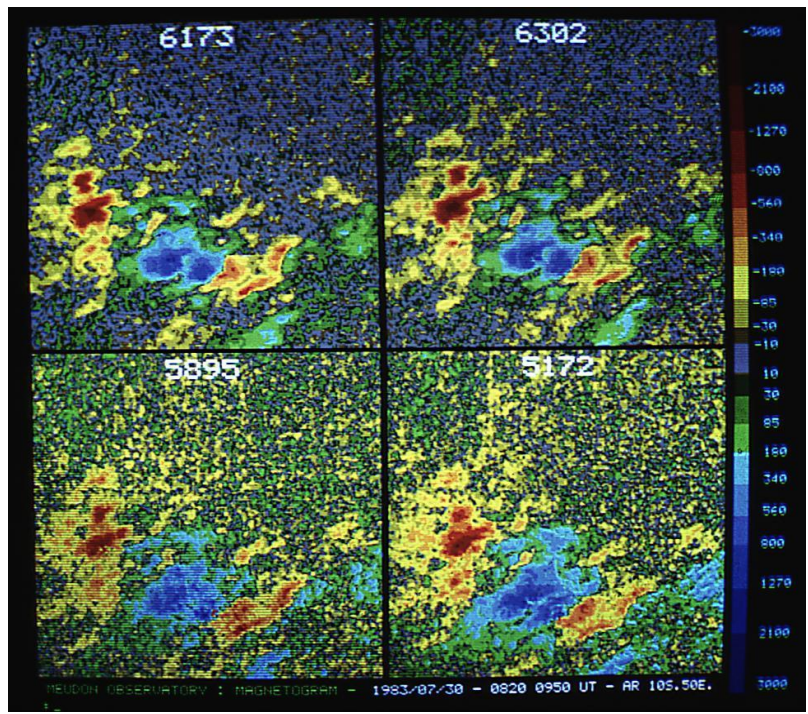

*Figure 14 : maps of magnetic fields in FeI 6173 Å, FeI 6302 Å, NaI 5895 Å, MgI 5172 Å, at different altitudes of the solar atmosphere, displayed in 1983 on the Meudon image processor. Courtesy OP.*

## IV - THE THEMIS MTR SPECTRO-POLARIMETER

Let us first introduce the THEMIS telescope. After the success of Meudon magnetograph, a much more powerful spectro-polarimeter was proposed to be built in a much better site than Meudon in terms of meteorological conditions, seeing and image quality.

The telescope should be polarization free in order to be able to measure the full Stokes vector without any contamination by parasitic effects. In 1980, the first project was a 60 cm aperture vacuum telescope and the suggested site was La Palma (Canary islands). Five years later, it became a 90 cm telescope in azimuthal mount with two vertical 8-metre spectrographs rotating with the telescope platform (Rayrole, 1985, and Figure 15). As it was extremely long to convince and get the necessary financial support, the telescope was inaugurated by king Juan Carlos in 1996 at Tenerife and commissioned in 1999.

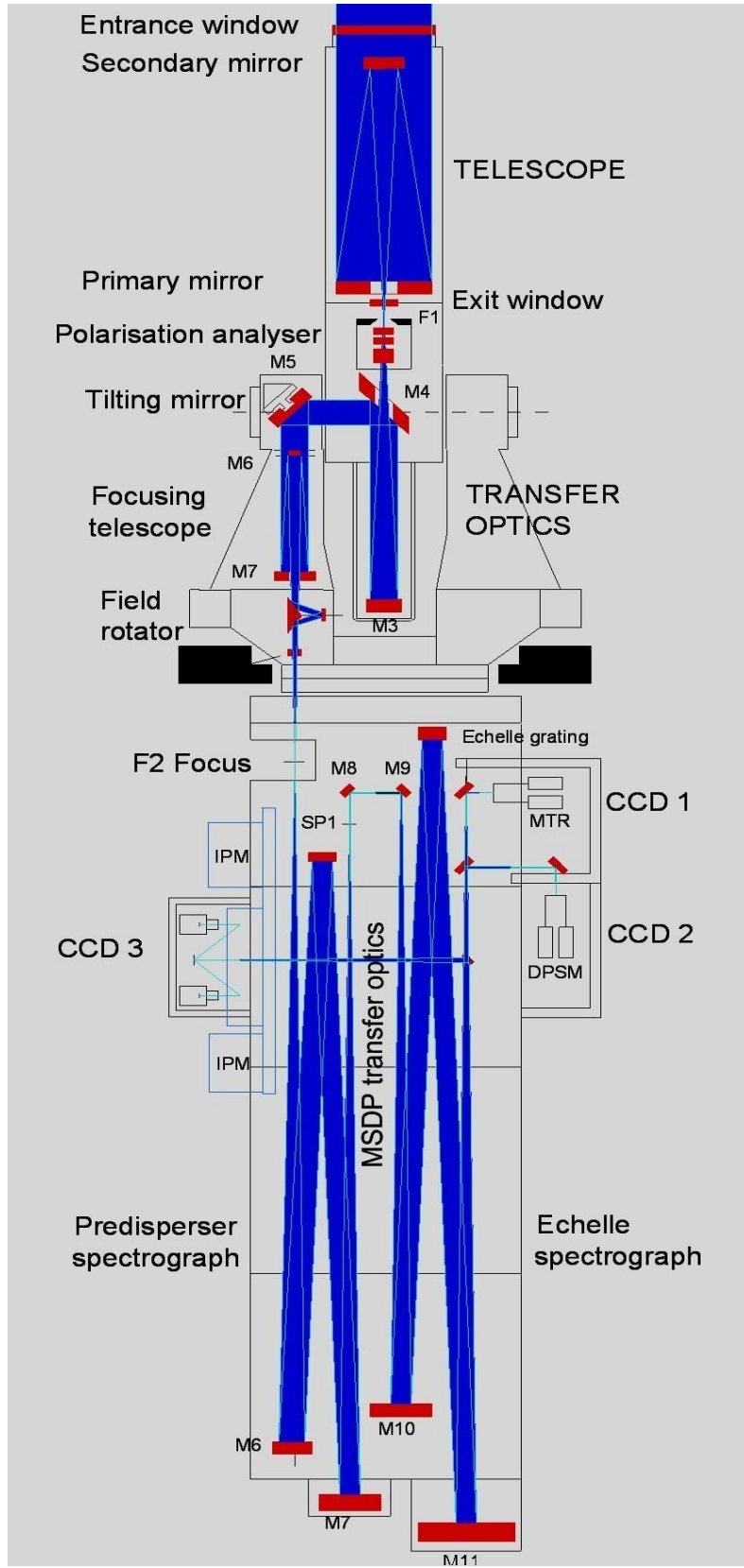

*Figure 15 : THEMIS is a polarization free solar telescope dedicated to spectro polarimetry, imaging polarimetry and vector magnetic field measurements. Courtesy OP.*

**The telescope**:

*Azimuthal mount, 90 cm aperture Ritchey Chrétien, Helium gas filled*

*F1 Focus (f = 17 m): polarimeter (two quarter waveplates + calcite birefringent beam splitter-shifter)*

**Transfert optics** *to the spectrographs:*

*Tip-tilt mirror M5*

*Small telescope M7, providing 3 X magnification*

*Field derotator (3 flat mirrors)*

**F2 focus** *(f = 57 m): 3 instruments*

**IPM** *(Italian Panoramic Monochromator), tunable Zeiss Lyot prefilter + Queensgate Fabry Pérot filter, 50 mÅ spectral resolution.*

*Two spectrographs (f = 8 m) including:*

**MTR** *mode (multi line spectroscopy), predisperser (8 m spectrograph 1, M6, M7, spectrum in SP1) and Echelle spectrograph (8 m spectrograph 2, 79 grooves/mm grating, 63° blaze angle, M10, M11, exit in CCD1), spectral resolution 20 mÅ.*

**MSDP** *mode (imaging spectroscopy, Multichannel Subtractive Double Pass): dispersion by spectrograph 1, beam shifter-splitter (16 channels) in SP1, dispersion subtraction by spectrograph 2, exit in CCD2), spectral resolution 80 mÅ.*

## IV – 1 – The THEMIS polarimeter

The polarimer (Figures 16 & 17), designed in the nineties, was made of two quarter waveplates (retardances $\delta_1 = \delta_2 = \pi/2$); the two waveplates were motorized and could rotate around the optical axis (Oz); their angles with Ox axis are $\alpha$ and $\beta$. The analyser included a birefringent polarizing beam splitter-shifter, which is equivalent to two polarizers with acceptance axis along Ox and Oy. With this device, couples of I+S and I-S signals could be measured simultaneously, with S = Q, U, V sequential (3 consecutive exposures); it was possible to practice beam exchange (Q → -Q, U → -U, V → -V), in order to improve the polarimetric precision (6 exposures required). Today the polarimeter uses liquid crystals.

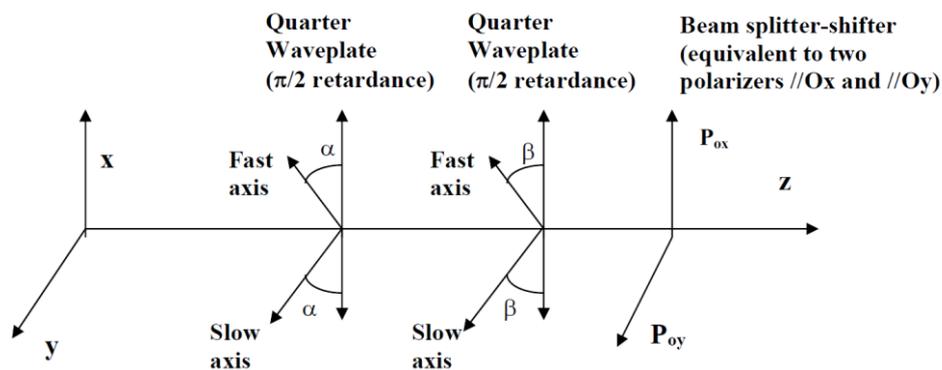

*Figure 16: the polarimeter principle ($\alpha$ and $\beta$ variable). Courtesy OP.*

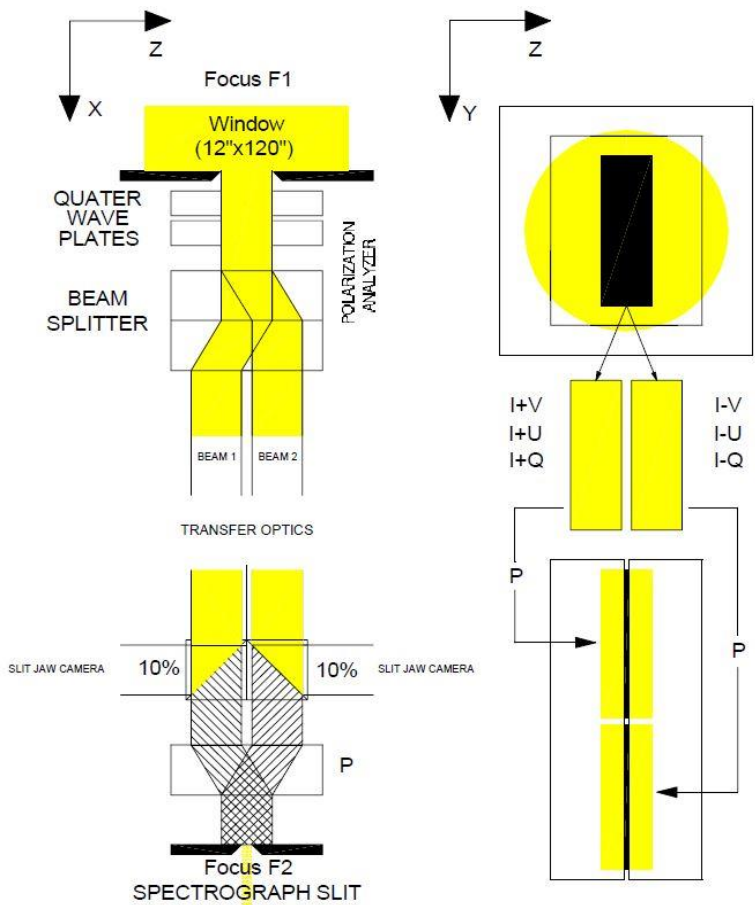

*Figure 17 : the dual beam arrangement before injection into the sepectrographs.*

***F1 focus*** *(f = 17 m, top):*

*A rectangular entrance window (12" x 120") is located at the F1 focus. The polarimeter delivers two shifted beams 1 and 2 delivering two simultaneous combinations of Stokes parameters I+S and I-S where S = Q, U, V and -Q, -U, -V in case of beam exchange.*

***F2 focus*** *(f = 57 m, bottom):*

*There is a slit jaw camera and the prisms (P) rearrange the two images of the entrance window to be aligned on the spectrograph slit before injection into the spectrograph.*

*Courtesy J. Rayrole and OP.*

With Müller matrices, one gets with **S**$_{in}$ and **S**$_{out}$ the input and output Stokes vectors:

**S**$_{out}$ = **P**$_{ox}$ **R**$_{-\beta}$ **T**$_{\pi/2}$ **R**$_{\beta}$ **R**$_{-\alpha}$ **T**$_{\pi/2}$ **R**$_{\alpha}$ **S**$_{in}$ for the Ox polarizer (beam 1), and at the same time, on beam 2,

**S**$_{out}$ = **P**$_{oy}$ **R**$_{-\beta}$ **T**$_{\pi/2}$ **R**$_{\beta}$ **R**$_{-\alpha}$ **T**$_{\pi/2}$ **R**$_{\alpha}$ **S**$_{in}$ for the Oy polarizer

The final result for the output intensity is:

$$I_{out} = \tfrac{1}{2} [ I_{in} \pm \{ Q_{in} ( \cos(2(\beta-\alpha)) \cos(2\alpha) \cos(2\beta) - \sin(2\alpha) \sin(2\beta) ) + U_{in} ( \cos(2(\beta-\alpha)) \sin(2\alpha) \cos(2\beta) + \cos(2\alpha) \sin(2\beta) ) + V_{in} \sin(2(\beta-\alpha)) \cos(2\beta) \} ]$$

The two signs correspond to the two polarizers (+ for Ox and – for Oy) or to the dual beam calcite polarizing beam shifter-splitter.

When $\alpha = \beta$, the two quarter waveplates are equivalent to a single half waveplate, so that:
$I_{out} = \tfrac{1}{2} [ I_{in} \pm \{ Q_{in} \cos(4\alpha) + U_{in} \sin(4\alpha) \} ]$
*In that case, one gets with 2 measures:*
*½ [I + Q] and ½ [I – Q] for $\alpha = \beta = 0$,*
*½ [I + U] and ½ [I – U] for $\alpha = \beta = \pi/8$*
*or with 4 measures for better precision with beam exchange:*
*½ [I + Q] and ½ [I – Q] for $\alpha = \beta = 0$,*
*½ [I – Q] and ½ [I + Q] (beam exchange) for $\alpha = \beta = \pi/4$,*
*½ [I + U] and ½ [I – U] for $\alpha = \beta = \pi/8$,*
*½ [I – U] and ½ [I + U] (beam exchange) for $\alpha = \beta = 3\pi/8$.*

When $\beta = 0$, $I_{out} = \tfrac{1}{2} [ I_{in} \pm \{ Q_{in} \cos^2(2\alpha) + U_{in} \cos(2\alpha) \sin(2\alpha) - V_{in} \sin(2\alpha) \} ]$
*In that case, one gets with a single measurement:*
*½ [I – V] and ½ [I + V] for $\alpha = \pi/4, \beta = 0$*
*or with 2 measures for better precision with beam exchange:*
*½ [I + V] and ½ [I – V] (beam exchange) for $\alpha = -\pi/4, \beta = 0$*

When $\alpha = 0$, $I_{out} = \tfrac{1}{2} [ I_{in} \pm \{ Q_{in} \cos^2(2\beta) + U_{in} \sin(2\beta) + V_{in} \sin(2\beta) \cos(2\beta) \} ]$
*In that case, one gets with a single measurement:*
*½ [I + U] and ½ [I – U] for $\alpha = 0, \beta = \pi/4$*
*or with 2 measures for better precision with beam exchange:*
*½ [I – U] and ½ [I + U] (beam exchange) for $\alpha = 0, \beta = -\pi/4$*

### IV – 2 – The MTR 2-camera mode (2 x 2', Figure 18)

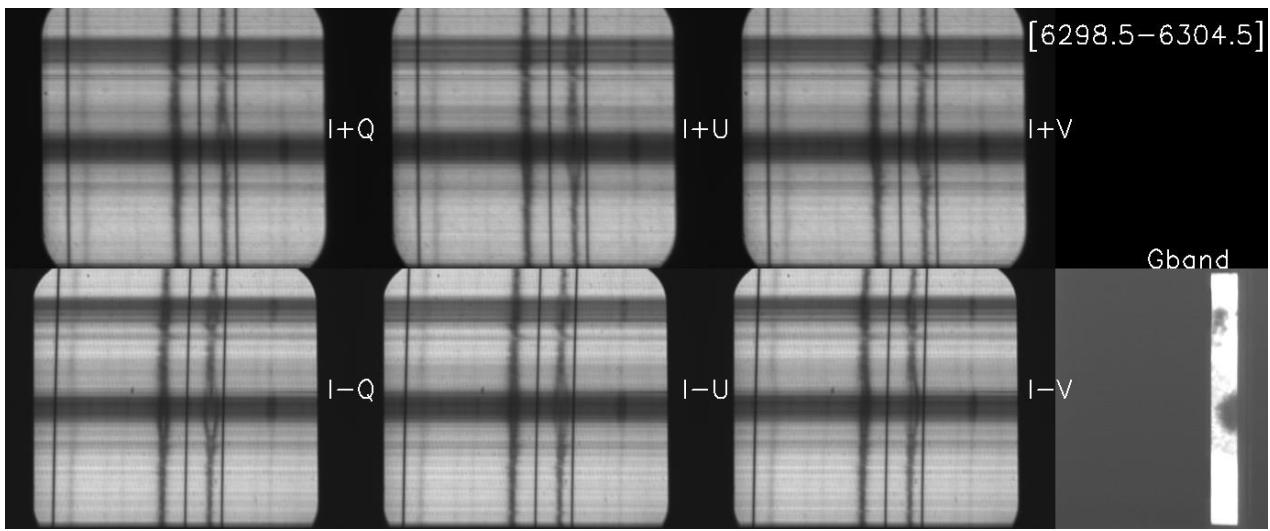

*Figure 18: Sunspots observations with the MTR 2-camera mode, FeI 6301.5 Å and 6302.5 Å lines and slit jaw in Gband. Camera 1 (top) delivers sequentially I+Q, I+U, I+V while camera 2 (bottom) observes I-Q, I-U, I-V. Couples I+S and I-S are simultaneous, with S = Q, U, V in sequence. Courtesy OP.*

The 2-camera mode (Figure 18) was the first offered in 1999 to the community. It used about 15 Thomson CCD cameras with 384 x 288 pixels of 20 µ (today new cameras are available). It allowed to derive the Stokes profiles and polarization rates, such as those of Figure 19.

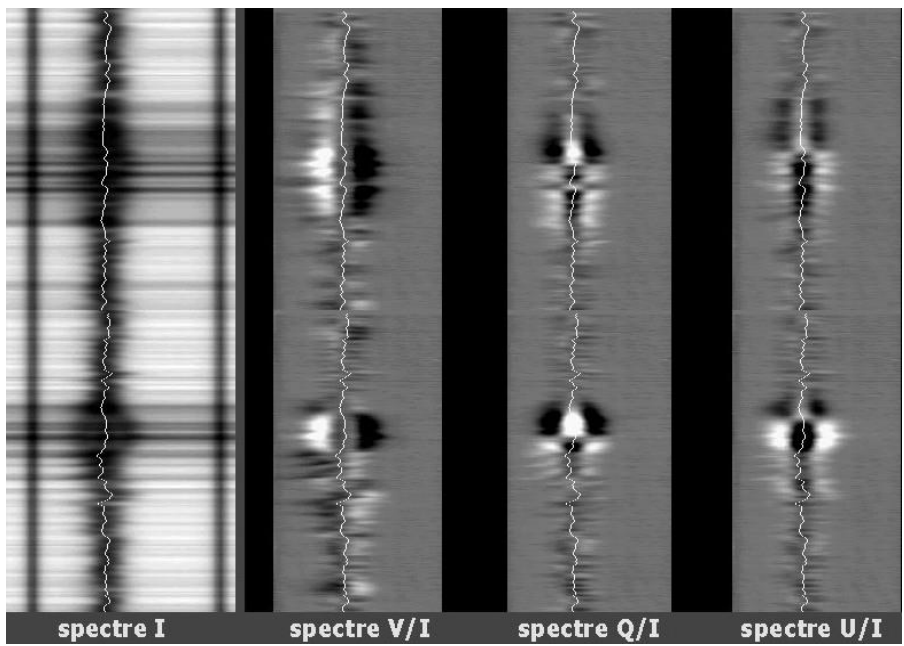

*Figure 19: Intensity I and polarization rate (V/I, Q/I, U/I) profiles of sunspots with the MTR 2-camera mode, FeI 6302.5 Å line. Courtesy J. Rayrole and OP.*

However, the 2-camera mode had some disadvantages, such as different CCD gain tables, flat field, optical path, line inclination, making the data processing complicated; some researchers tried the 1-camera mode in order to improve the polarimetric precision.

**IV – 3 –The MTR 1-camera mode (2 x 1', Figure 20)**

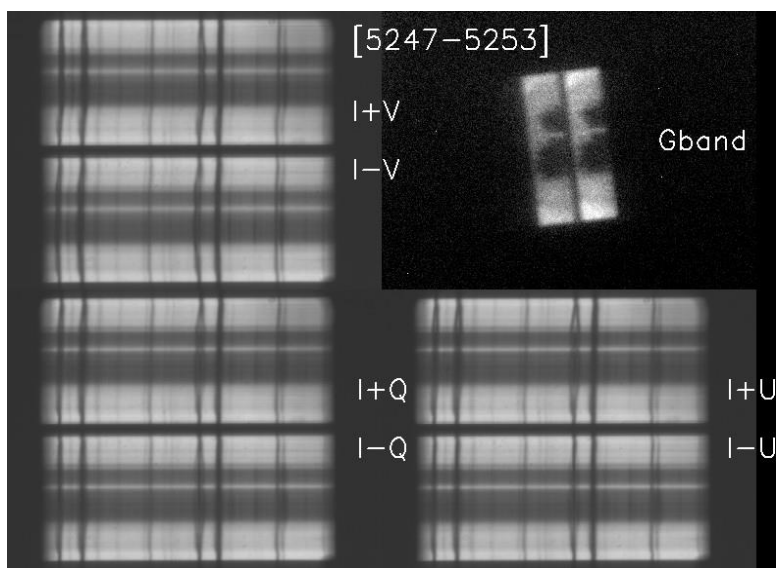

*Figure 20: Sunspots observations with the MTR 1-camera mode (2 x 1'), FeI lines around 5250 Å and slit jaw in Gband. The unique camera delivers simultaneous couples I+V and I-V, then I+Q and I-Q, and finally I+U and I-U (3 exposures are needed). Courtesy OP.*

The 1-camera mode had the disadvantage of reducing the FOV, but the big advantage of the unique CCD gain table, flat field, closer optical path and line inclination, improving the polarimetric precision. The ultimate step was the beam exchange introduced by Meir Semel (1932-2012).

## IV – 4 –The 1-camera mode with beam exchange (2 x 1', Figure 21)

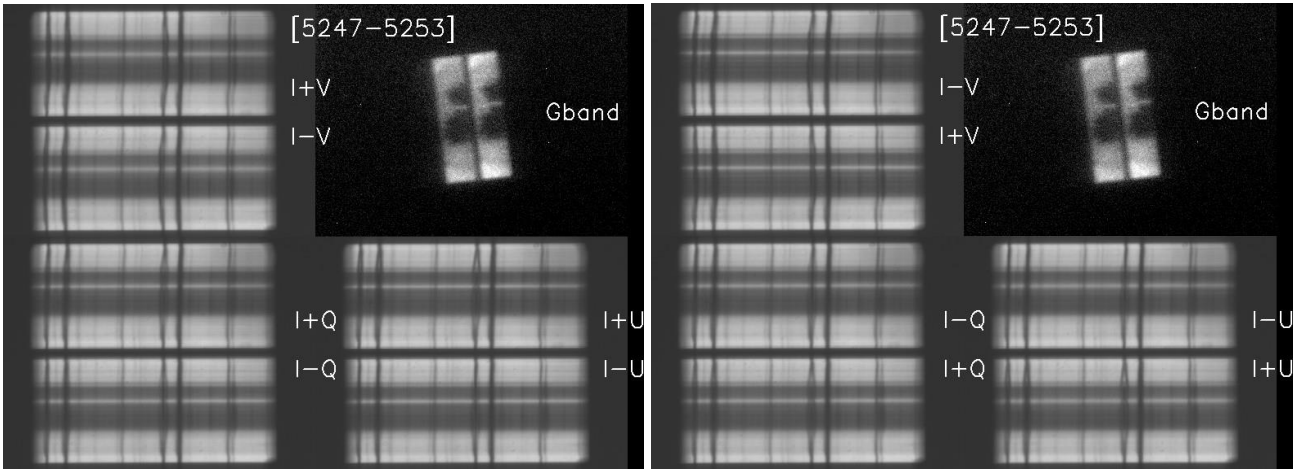

*Figure 21: Sunspots observations with the MTR 1-camera mode (2 x 1'), FeI lines around 5250 Å and slit jaw in Gband with beam exchange. The camera delivers simultaneous couples I+V and I-V, then I-V and I+V (beam exchange), the same for parameters Q and U, so that 6 exposures are needed. Courtesy OP.*

The beam exchange introduced by Semel et al (1993), which consists to record on the same CCD area alternatively the I+S and I-S signals, where S = (Q, U, V), has the big advantage to be independent of the CCD gain table and flat field, and therefore considerably improved the polarimetric precision, but only when the 6 exposures benefit of efficient image stabilization by the tip tilt (or today adaptive optics).

## IV – 5 – The weak field approximation to derive the magnetic field vector

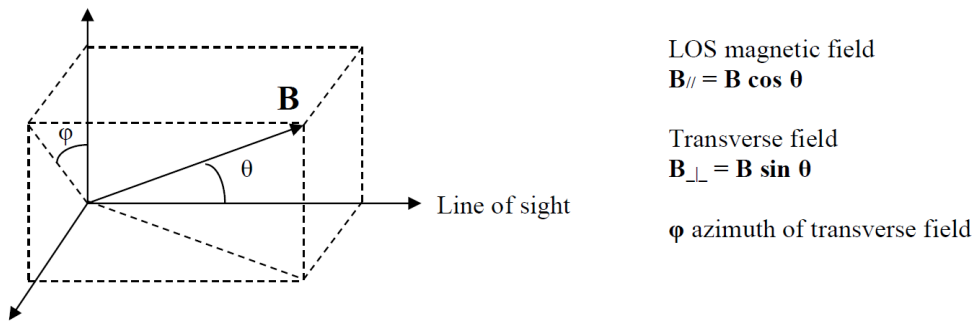

LOS magnetic field
$B_{//} = B \cos \theta$

Transverse field
$B_\perp = B \sin \theta$

$\varphi$ azimuth of transverse field

*Figure 22: the components of the magnetic field vector (LOS and transverse components, plus azimuth)*

In the weak field approximation:

$V(\lambda) = \Delta\lambda_{B_{//}} \, dI/d\lambda$  where $\Delta\lambda_{B_{//}} = 4.67 \times 10^{-13} B_{//} g^* \lambda^2$ is proportional to $B_{//}$

$Q(\lambda) = -1/4 \, \Delta\lambda_{B_\perp}^2 \cos(2\varphi) \, d^2I/d\lambda^2$  and

$U(\lambda) = -1/4 \, \Delta\lambda_{B_\perp}^2 \sin(2\varphi) \, d^2I/d\lambda^2$  where $\Delta\lambda_{B_\perp} = 4.67 \times 10^{-13} B_\perp g^* \lambda^2$ is proportional to $B_\perp$

$(Q^2 + U^2)^{1/2} = 1/4 \, \Delta\lambda_{B_\perp}^2 \, |d^2I/d\lambda^2|$ gives the strength of the transverse field $B_\perp$

$U/Q = \tan(2\varphi)$ provides $\varphi$ (modulo $\pi$)

Q and U are second orders in B, V is first order. In these formulae, B is expressed in Gauss (1G = $10^{-4}$ T) and $\lambda$ in Å. The derivatives $dI/d\lambda$ et $d^2I/d\lambda^2$ are without any magnetic perturbation. Stokes measures require high resolution spectroscopy. It is not possible to get V from integration through a broad band coloured filter, but it is possible to estimate Q et U when the filter includes many magnetic lines, such as in the blue part of the spectrum (Leroy, 1962). When the unperturbed profile has a gaussian shape, such as:

$I(\lambda) = I_c ( 1 - r\, e^{-1/2\, ((\lambda-\lambda_0)/\Delta\lambda)^2} )$

At line centre ($\lambda = \lambda_0$), $V/I(\lambda_0) = 0$. But at inflexion points ($\lambda = \lambda_0 \pm \Delta\lambda$), the circular polarization rate is:

$V/I\,(\lambda_0 \pm \Delta\lambda) = \pm (\Delta\lambda_{B//}/\Delta\lambda)\, [r\, e^{-1/2}/(1 - r\, e^{-1/2})]$

On the contrary, at line centre ($\lambda = \lambda_0$), one gets:

$Q/I(\lambda_0) = -1/4\, (\Delta\lambda_{B\_\bot\_}/\Delta\lambda)^2\, \cos(2\phi)\, [r/(1-r)]$ and

$U/I(\lambda_0) = -1/4\, (\Delta\lambda_{B\_\bot\_}/\Delta\lambda)^2\, \sin(2\phi)\, [r/(1-r)]$

Linear polarization rate is: $(Q^2 + U^2)^{1/2}/I(\lambda_0) = 1/4\, (\Delta\lambda_{B\_\bot\_}/\Delta\lambda)^2\, [r/(1-r)]$

The circular and linear polarization rates are respectively proportional to $(\Delta\lambda_{B//}/\Delta\lambda)$ and $(\Delta\lambda_{B\_\bot\_}/\Delta\lambda)^2$. An example of polarization rates is displayed in Figure 23.

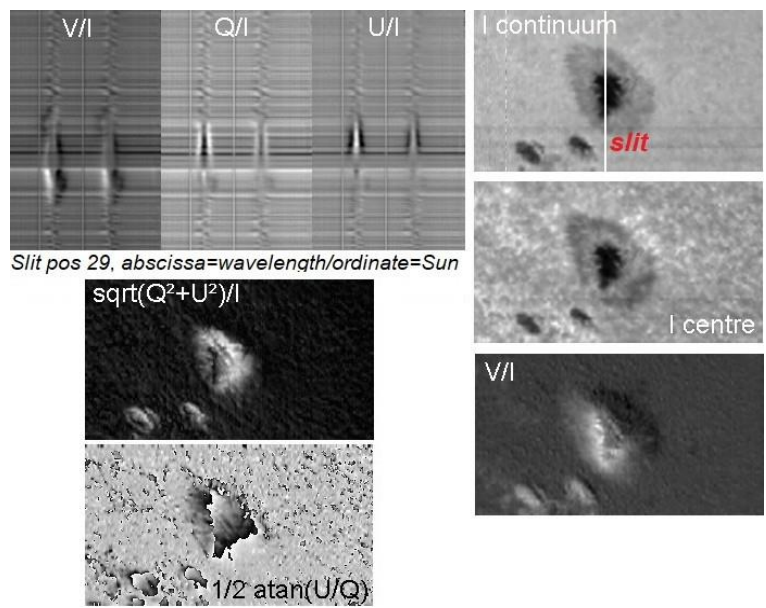

*Figure 23: polarization rates derived from the spectro-polarimetry of a sunspot. Courtesy OP.*

**IV – 6 –The MTR 1-camera mode with grid (2 x 1', Figure 24)**

In order to observe simultaneously I+S and I-S signals (S = Q, U, V) in optical conditions as close as possible, which is required for weak polarizations, Semel (1980) introduced a grid in the solar image, forming the dual beams 1-1', 2-2', 3-3'… of Figure 24. This system was implemented at the F1 focus.

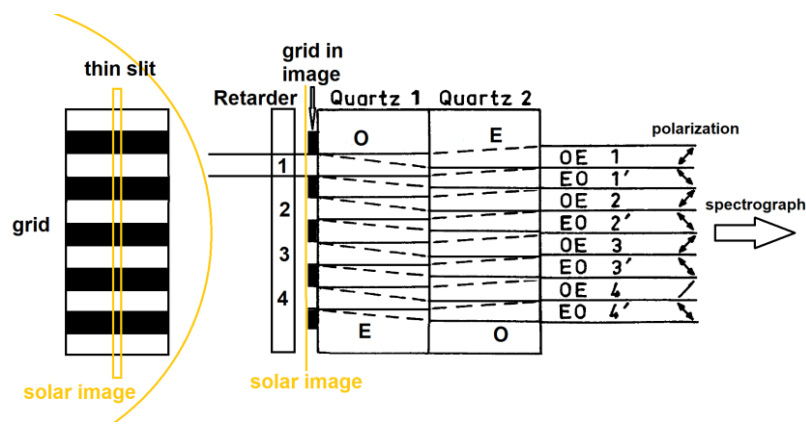

*Figure 24: Semel's grid is located in the solar image at the F1 focus in front of the polarimeter, which forms optically close dual beams 1-1', 2-2'… before injection into the spectrograph. Courtesy OP.*

Figures 25 shows an exemple of observations with Semel's grid. The grid step was about 15". Three exposures are needed in full Stokes polarimetry, and even six in the case of beam exchange (Figure 26) when one requires a better precision in the case of weak polarization measurements, such as lines of the second solar spectrum near the limb (Figure 27).

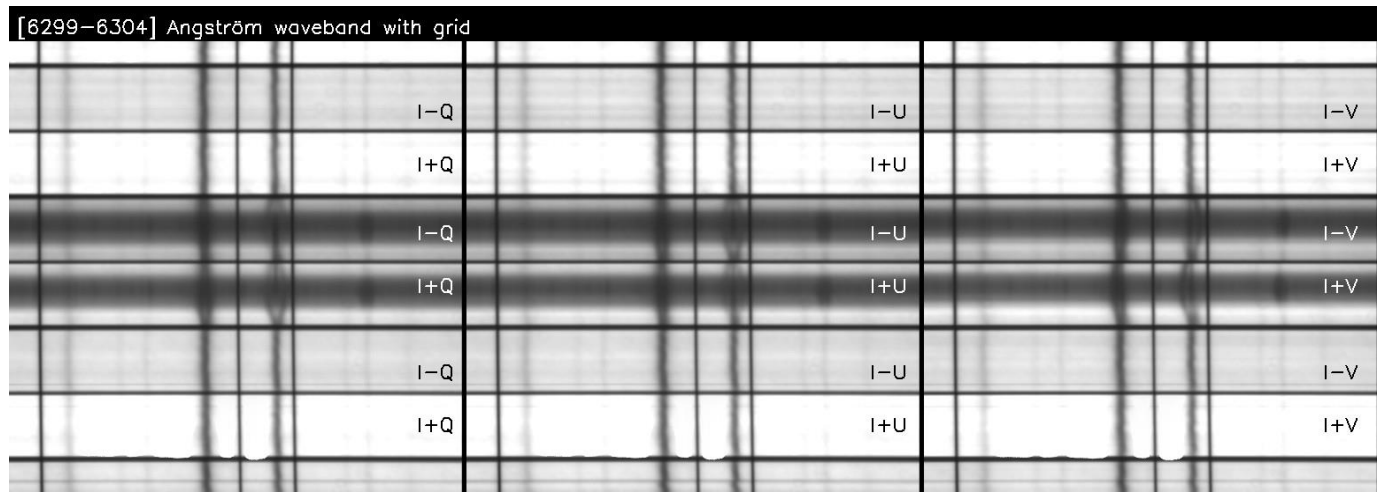

*Figure 25: three successive exposures of the camera over a sunspot with FeI 6301.5 Å and 6302.5 Å lines with Semel's grid at the F1 focus. Exposure 1 produces alternate strips of I-Q, I+Q, exposure 2 I-U and I+U and exposure 3 I-V and I+V. Of course, half of the FOV is masked by this method. Courtesy OP.*

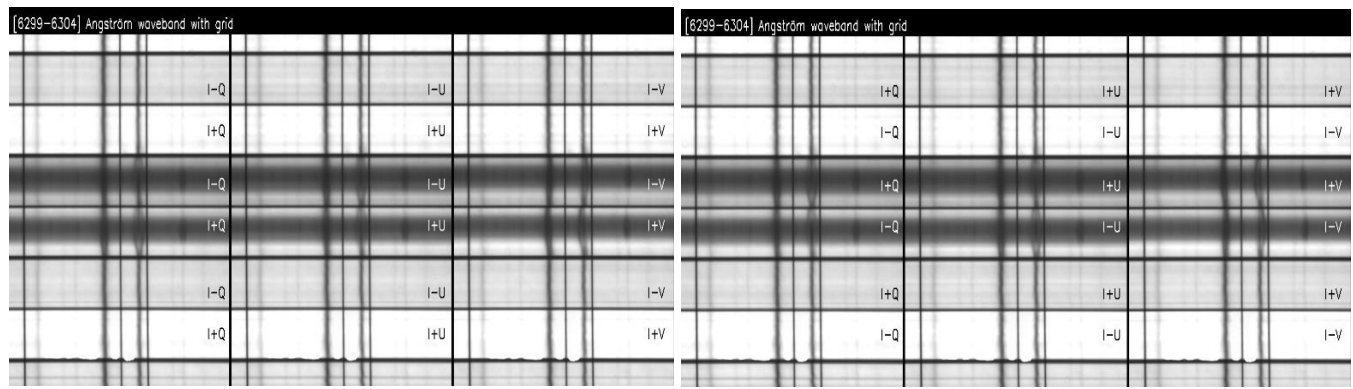

*Figure 26: six successive exposures of the camera over a sunspot with FeI 6301.5 Å and 6302.5 Å lines with Semel's grid at the F1 focus. Alternate strips of I-Q, I+Q (exp 1), I-U, I+U (exp 2), I-V, I+V (exp 3), beam exchange I+Q, I-Q (exp 4), I+U, I-U (exp 5), I+V, I-V (exp 6) are recorded. Courtesy OP.*

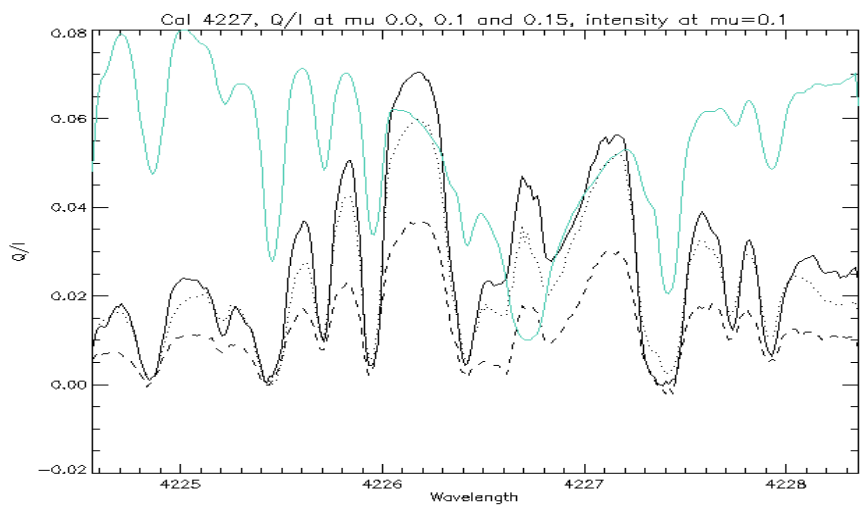

*Figure 27: weak field polarization Q/I of CaI 4227Å near the limb at µ=0.0, 0.1 and 0.15. Courtesy OP.*

# V - POLARIMETRY WITH LIQUID CRYSTALS AT THE TURRET DOME

With two waveplates of variable retardance (such as liquid crystals, Figure 28) but fixed azimuth (retardance $\delta 1$, azimuth 0, and retardance $\delta 2$, azimuth $\pi/4$), it is possible to measure simultaneously couples of I+S and I-S signals (where S = Q, U, V in sequence); beam exchange is even possible with 6 consecutive exposures (instead of 3) to get the Stokes vector at higher precision. The polarimeter of Figure 29 started operations in 2003 at the spectrograph of the Pic du Midi Turret Dome (Roudier, 2022).

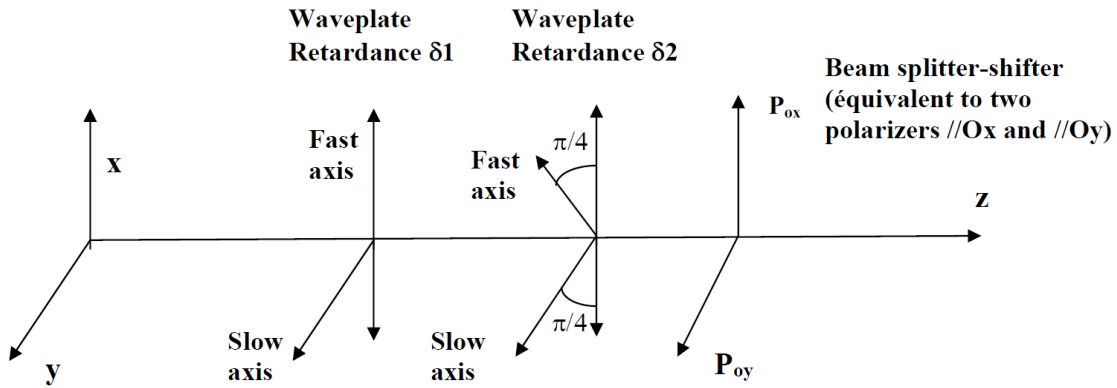

*Figure 28: principle of liquid crystal polarimeters for full Stokes measurements. Courtesy OP.*

With Müller matrices, one gets with **S**in and **S**out the input and output Stokes vectors:
**S**out = **P**ox **R**–π/4 **T**δ2 **R**π/4 **T**δ1 **S**in with the polariser //Ox
And simultaneously:
**S**out = **P**oy **R**–π/4 **T**δ2 **R**π/4 **T**δ1 **S**in with the polariser //Oy
The output intnsity is:

| $I_{out} = ½ [ I_{in} ± \{ Q_{in} \cos(\delta 2) + \sin(\delta 2) ( U_{in} \sin(\delta 1) – V_{in} \cos(\delta 1) ) \} ]$ |
|---|

The two signs correspond to the two polarizers (+ for Ox and – for Oy) or to the dual beam polarizing beam shifter-splitter.

| Retardance | Single beam<br>Polariser //Ox | Dual beam<br>Birefringent beam-shifter/splitter | |
|---|---|---|---|
| $\delta 1 = 0, \delta 2 = 0$, | $I_{out} = ½ [ I_{in} + Q_{in} ]$ | $I_{out} = ½ [ I_{in} + Q_{in} ] , ½ [ I_{in} - Q_{in} ]$ | |
| $\delta 1 = 0, \delta 2 = \pi$, | $I_{out} = ½ [ I_{in} - Q_{in} ]$ | $I_{out} = ½ [ I_{in} - Q_{in} ] , ½ [ I_{in} + Q_{in} ]$ | beam exchange |
| $\delta 1 = 0, \delta 2 = \pi/2$, | $I_{out} = ½ [ I_{in} - V_{in} ]$ | $I_{out} = ½ [ I_{in} - V_{in} ] , ½ [ I_{in} + V_{in} ]$ | |
| $\delta 1 = 0, \delta 2 = 3\pi/2$, | $I_{out} = ½ [ I_{in} + V_{in} ]$ | $I_{out} = ½ [ I_{in} + V_{in} ] , ½ [ I_{in} - V_{in} ]$ | beam exchange |
| $\delta 1 = \pi/2, \delta 2 = \pi/2$, | $I_{out} = ½ [ I_{in} + U_{in} ]$ | $I_{out} = ½ [ I_{in} + U_{in} ] , ½ [ I_{in} - U_{in} ]$ | |
| $\delta 1 = \pi/2, \delta 2 = 3\pi/2$, | $I_{out} = ½ [ I_{in} - U_{in} ]$ | $I_{out} = ½ [ I_{in} - U_{in} ] , ½ [ I_{in} + U_{in} ]$ | beam exchange |

In case of a single beam, single camera, when the retardances $\delta 1, \delta 2$ vary at high frequency (1000 Hz), the seeing (100 Hz) is frozen; this is the principle of the Zürich Imaging POLarimeter (ZIMPOL), one of the most famous polarimeters. At Pic du Midi, we first used a ferro electric half waveplate (FLC from Displaytech company, able to support kHz modulation) and nematic crystals (from Meadowlark) with variable retardance, but limited at 30 Hz (Malherbe *et al*, 2007a, 2007b). As our camera was commercial and limited to 10 Hz, we got with our modulator induced seeing cross talk, contrarily to ZIMPOL which works with a kHz custom camera. Some observations are displayed on Figure 30 (full Stokes polarimetry of sunspots) and 31 (second solar spectrum in imagery mode and in spectroscopic mode).

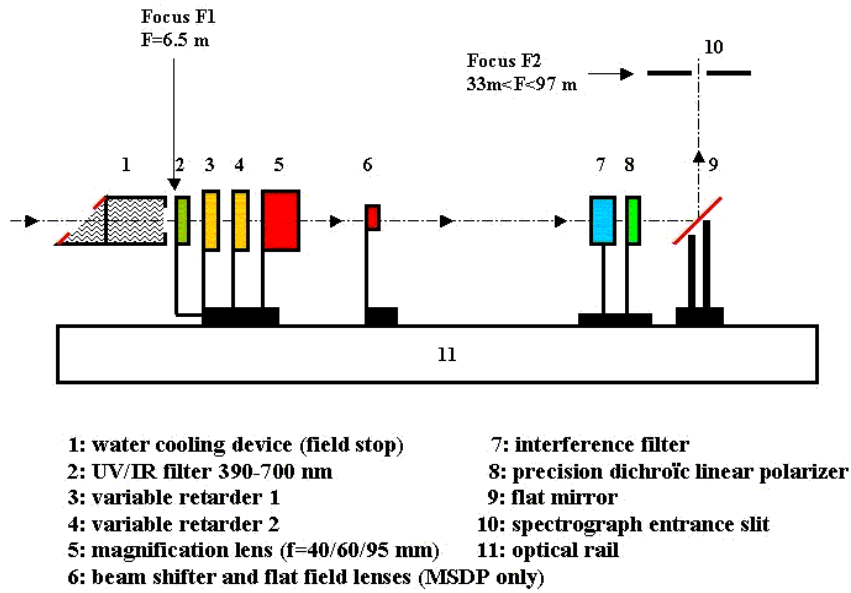

1: water cooling device (field stop)
2: UV/IR filter 390-700 nm
3: variable retarder 1
4: variable retarder 2
5: magnification lens (f=40/60/95 mm)
6: beam shifter and flat field lenses (MSDP only)
7: interference filter
8: precision dichroïc linear polarizer
9: flat mirror
10: spectrograph entrance slit
11: optical rail

*Figure 29: Pic du Midi Turret Dome single beam polarimeter in 2003. Courtesy OP.*

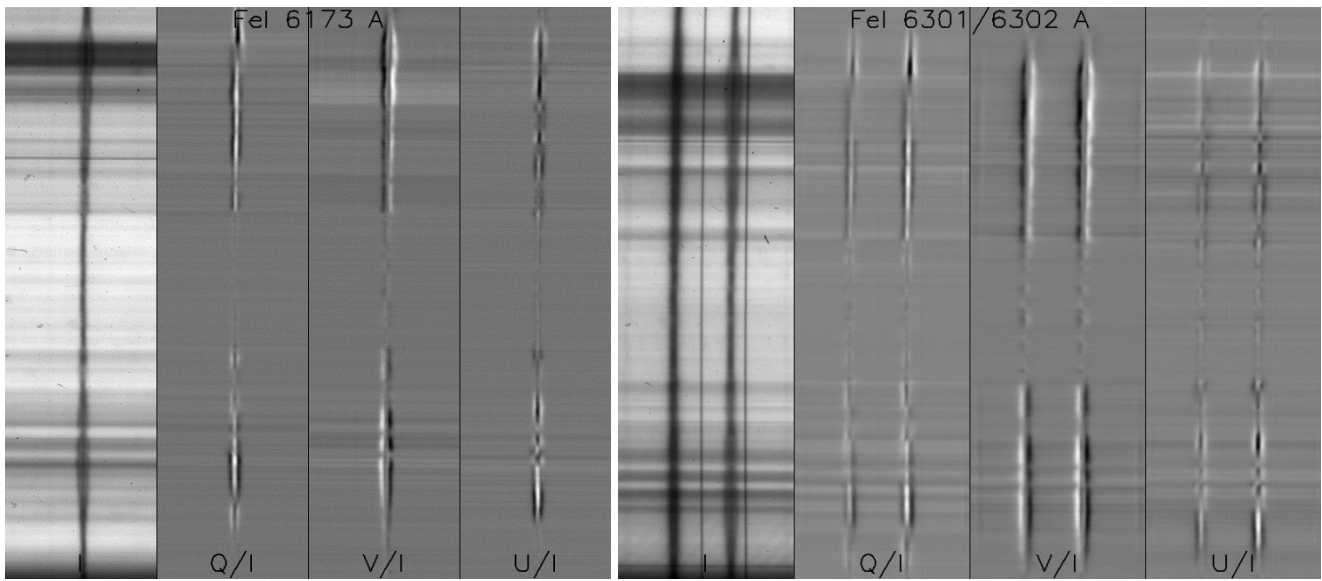

*Figure 30: Full Stokes polarimetry of sunspots in FeI 6173 Å and FeI 6301/6302 Å with the liquid crystal polarimeter and spectrograph of the Pic du Midi Turret Dome. Courtesy OP.*

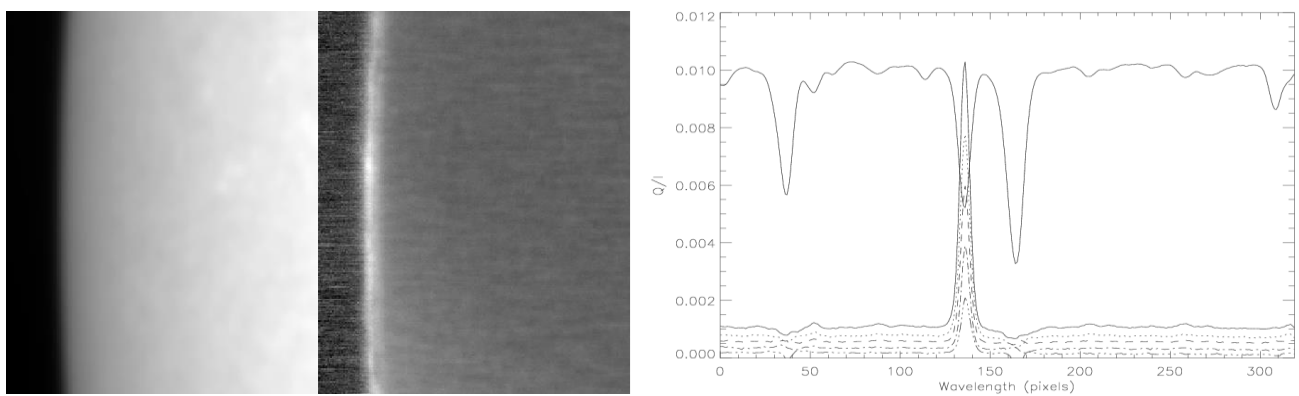

*Figure 31: linear polarization at the limb in imagery mode (460 nm, 10 nm FWHM, intensity and Q/I, left) and spectroscopy of SrI 4607 Å at various distances of the solar limb with the liquid crystal polarimeter of the Pic du Midi Turret Dome. Courtesy OP.*

We had a project of a dual beam full Stokes polarimeter with a new beam shifter-splitter designed by Semel (Figure 32), somewhat similar to the separator previously built for the THEMIS polarimeter (Figure 17); unfortunately, this high precision optical system took a long time and arrived too late, because the retirement of many people inhibited to adjust and test it carefully, so that it was never used.

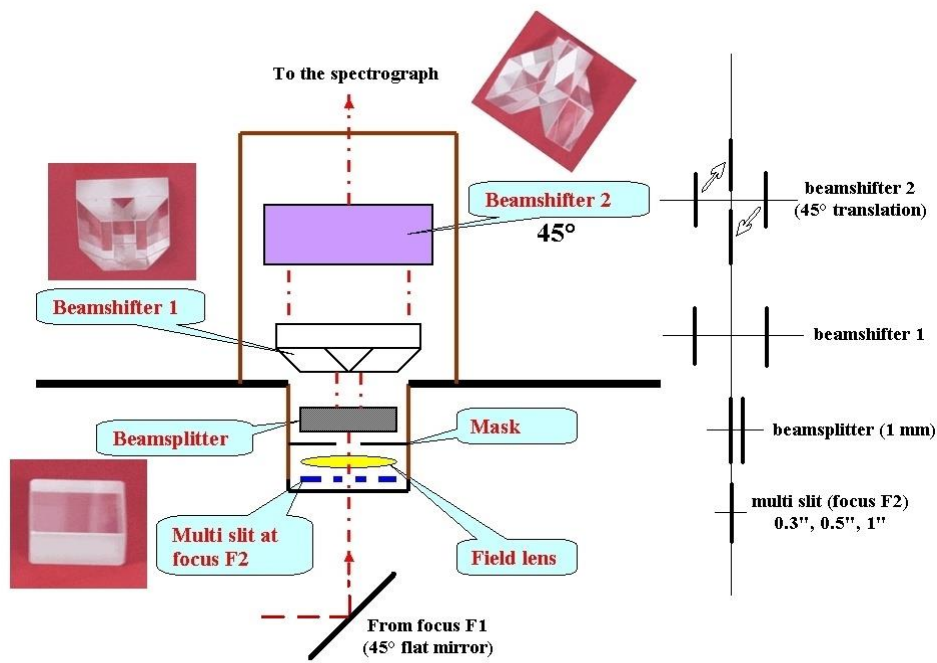

*Figure 32: the beam shifter-splitter for the Pic du Midi Turret Dome polarimeter. Courtesy OP.*

## VI - POLARIMETRY WITH THE MSDP

The Multichannel Subtractive Double Pass (MSDP, Figure 33) is an imaging spectroscopy or spectro polarimetric mode, which was implemented in many spectrographs, such as those of Meudon Solar Tower (1977), the Pic du Midi Turret Dome (1980), the Vacuum Tower Telescope in Tenerife (1990), THEMIS (1996) and the Bialkow coronagraph (Poland). Let us first explain the principle of the MSDP (Mein, 1977, Mein *et al*, 2021) based on Figure 33. The thin slit of the spectrograph is replaced by a rectangular window (F1) which provides a 2D FOV. C1 and C1 are collimator and chamber mirrors. F2 is the spectrum focus. In F2, there is a slicer and beam shifter delivering N channels, detailed in Figure 34.

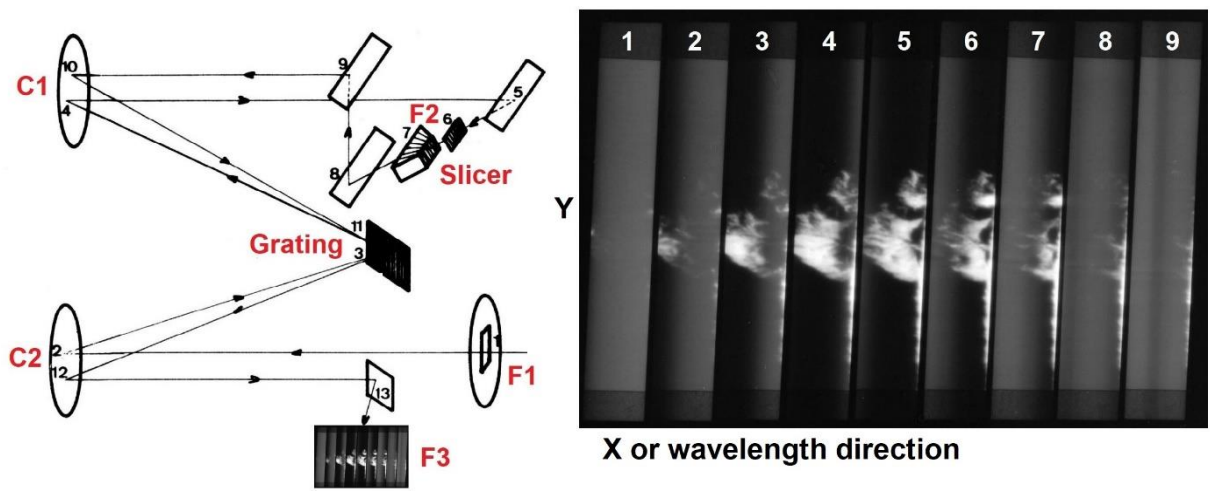

*Figure 33: principle of MSDP imaging spectroscopy with a slicer in the spectrum at F2. Courtesy OP.*

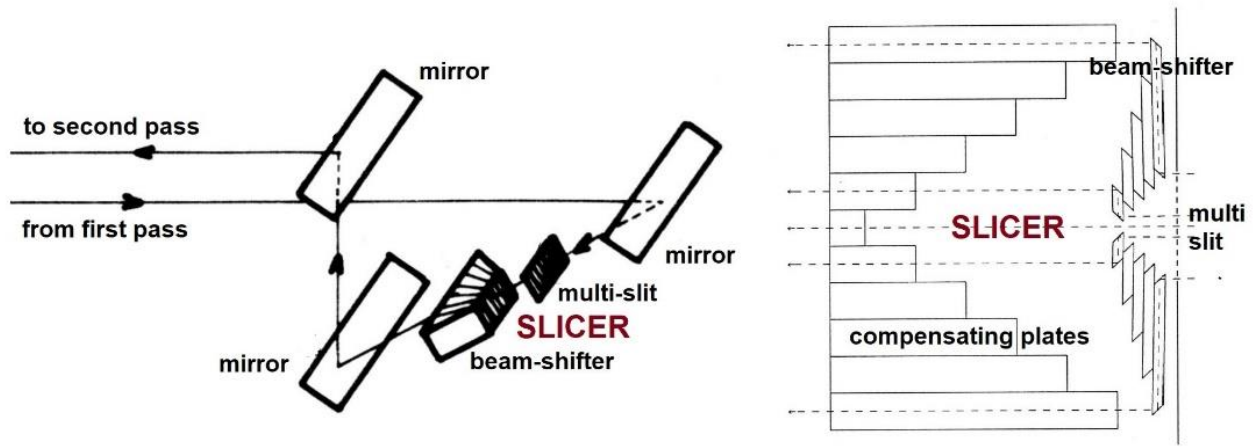

*Figure 34: the MSDP slicer in the spectrum focus delivering N channels. N was progressively extended at Meudon from 7 to 18 with various technologies (multi-slit in the spectrum or micro-mirrors); N = 16 at THEMIS; N = 24 for the future Solar Line Emission Dopplerometer. Courtesy OP.*

The beam is reinjected into the spectrograph for a subtractive second pass on the grating and forms in F3 a spectra image, composed of N channels covering a spectral line. The spatial and spectral information are mixed, as the wavelength varies linearly from the left to the right of each channel. There is a global wavelength shift Δλ between channel n to channel n+1. For instance, the FOV at Meudon or at the VTT Tenerife is 1' x 8' and the shift Δλ between two consecutive channels is 0.30 Å in Hα (Figure 35). At THEMIS, we had Δλ = 0.08 Å and recently Δλ = 0.04 Å was achieved with a new prototype at Meudon.

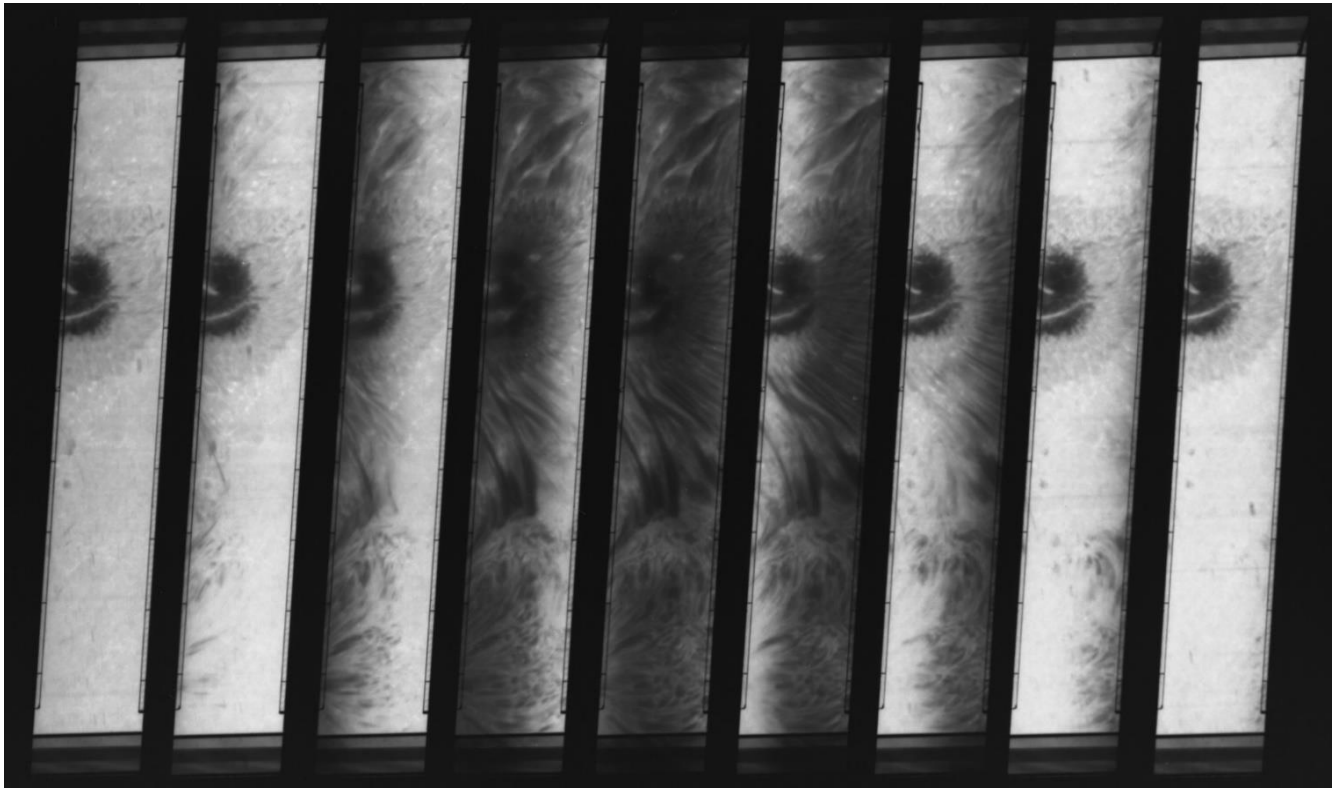

*Figure 35: the MSDP of the VTT at Tenerife in Hα. The wavelength shift between two consecutive channels is Δλ = 0.3 Å, it is governed by the step s (in mm) of the multi-slit of the slicer. At abscissa x inside channel number n, the wavelength is $\lambda(x) = \lambda_0 + x/d + n \, \Delta\lambda$ where d is the dispersion of the spectrograph (mm/Å) and Δλ = s/d. If w is the width of the rectangular entrance window (in mm), the wavelength variation inside each channel is w/d. Hence, channels are not monochromatic. Courtesy OP.*

## VI – 1 – Single beam polarimetry with liquid crystals and image selection

High spatial observations of the solar granulation and sunspots were got at the MSDP of the Pic du Midi Turret Dome, as shown by Figure 36, in NaD1 5896 Å line. The polarimetric strategy was modulation combined to fast data acquisition and short exposure time with a commercial CCD camera (5-10 Hz). Image stabilization was not available, and this moderate cadence did not allow to freeze the atmospheric turbulence (100 Hz), so that we used instead image selection, destretching and summing magnetic maps (Figure 37) to reduce the noise (Malherbe *et al*, 2004, Roudier *et al*, 2006).

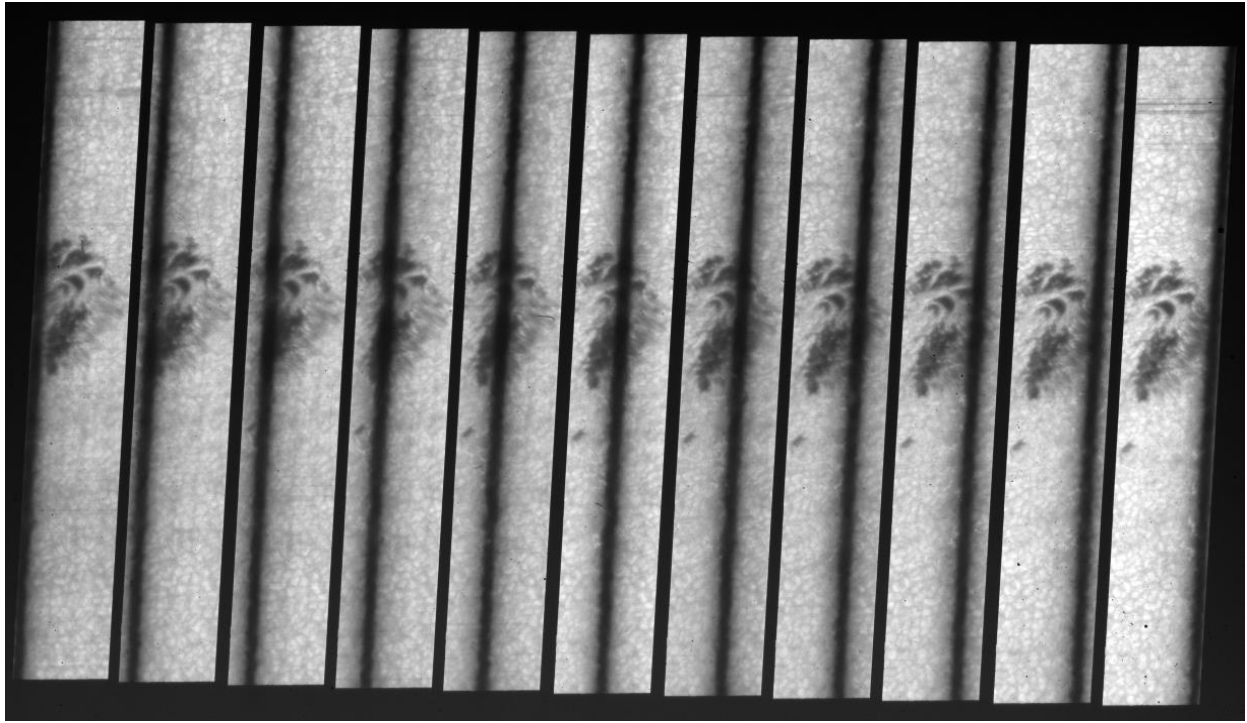

*Figure 36: the MSDP at Pic du Midi in NaD1 line. Courtesy OP.*

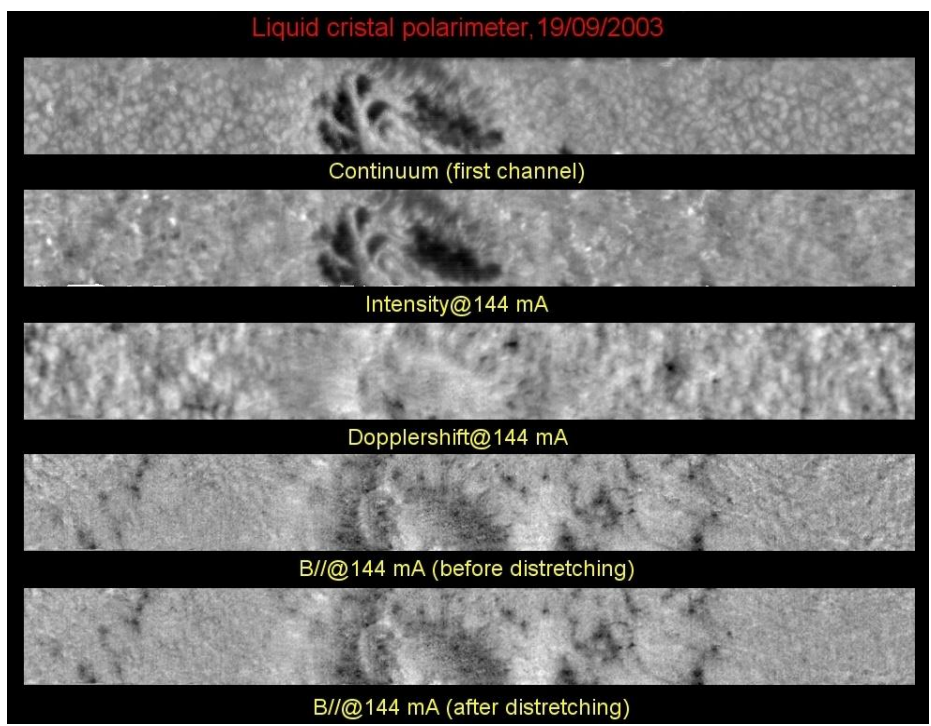

*Figure 37: NaD1 magnetometry at Pic du Midi; from top to bottom, continuum intensity, line intensity, Dopplershifts, LOS magnetic field before and after image destretching. Courtesy OP.*

## VI – 2 – Dual beam polarimetry with the MSDP: two strategies

We developed two strategies for dual beam MSDP polarimetry, as shown by Figure 38. The grid allows beam splitting in the Y direction (THEMIS method), while half image masking implies beam splitting in the X direction (Meudon method). Both have advantages and disadvantages. The grid delivers many segmented small FOVs (complex spatial reconstruction), but I+S and I-S signals (S = Q, U, V) have the same wavelength in the channels. On the contrary, half image masking has a unique FOV, but I+S and I-S are not measured at the same wavelengths in the channels (complex spectral reconstruction).

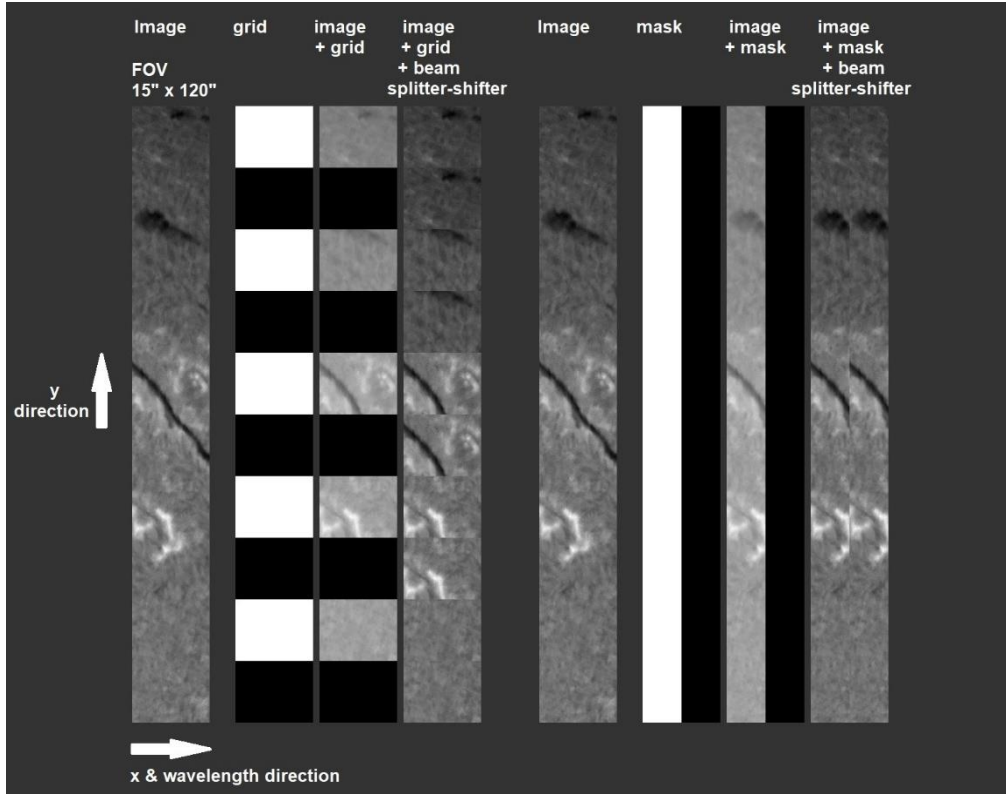

*Figure 38: two strategies for MSDP polarimetry. Left: grid and beam splitting in the spatial Y direction. Right: half image masking and beam splitting in the spatial X (or wavelength) direction. Courtesy OP.*

## VI – 3 – Dual beam polarimetry with the grid at the MSDP THEMIS

The grid method was implemented at THEMIS with N = 16 channels (Figure 39) and full Stokes polarimetry allowed Mein *et al* (2009) to derive the vector magnetic field of sunspots (Figure 40).

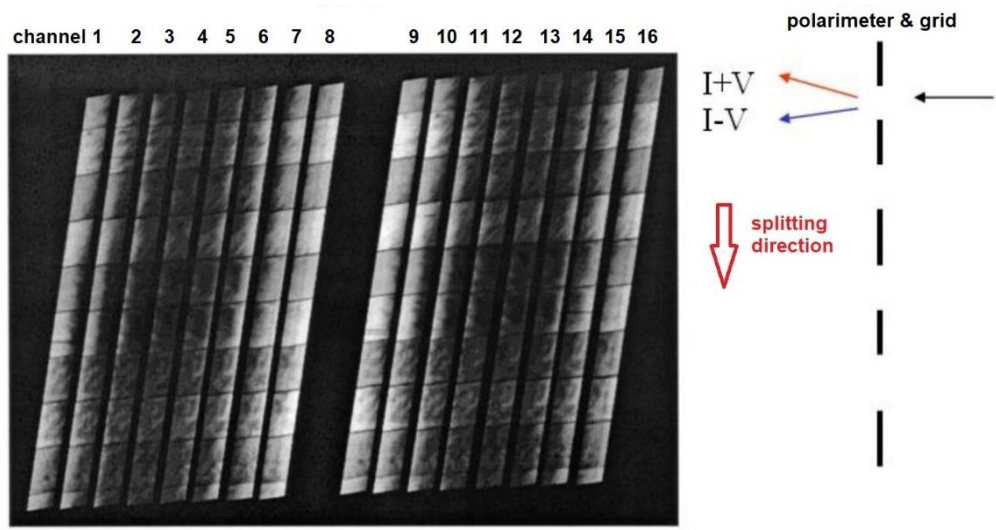

*Figure 39: dual beam polarimetry with grid at the MSDP THEMIS*

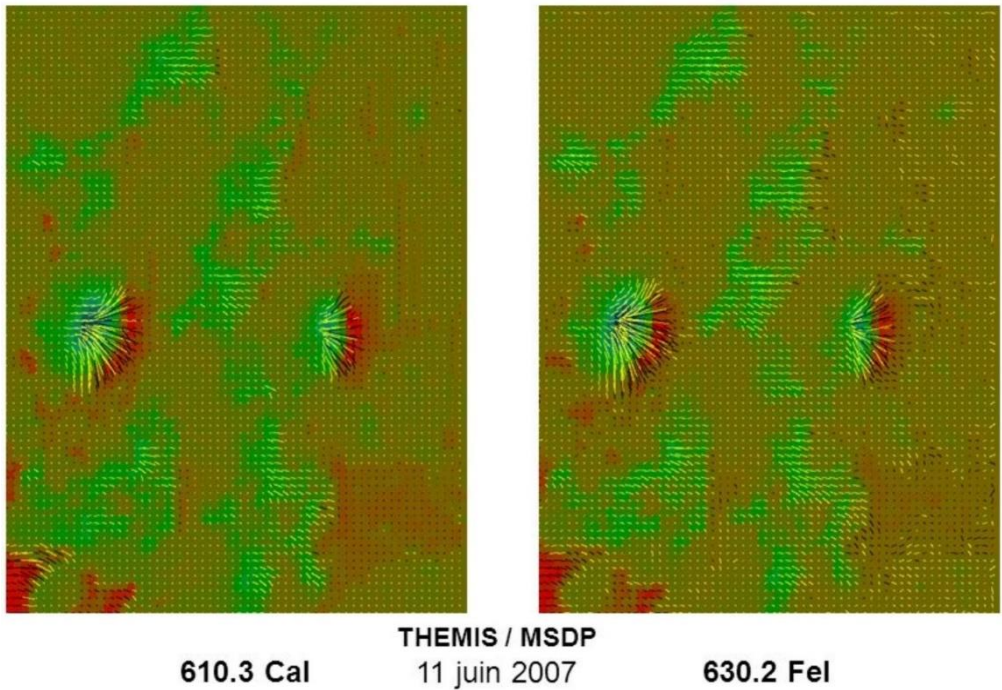

*Figure 40: vector magnetic field reconstruction above a sunspot in CaI 6103 Å and FeI 6302 Å with the MSDP THEMIS in dual beam polarimetry with grid. After Mein et al, 2009.*

**VI – 4 – Dual beam polarimetry with image masking at Meudon**

We used the liquid crystal polarimeter of Figure 41 with a calcite beam splitter-shifter. The entrance window is reduced in size in X direction by a factor 2 in comparison with no polarimetry (half image masking technique). Dual beam full stokes measurements can be achieved with modulation. Due to instrumental polarization of the 2-mirror coelostat and a 45° Newton mirror, we restricted us to circular polarization (I+V, I-V) observations at the cadence permitted by our old CCD camera (0.5 Hz). The MSDP spectra image in MgI 5173 Å of Figure 42 shows that each of the 18 available channels (formed by the new technology micro-mirror slicer) are divided into 2 subchannels I+V, I-V by half image masking, so that we have 36 subchannels of the same FOV. However, for a given (x, y) point of the Sun, identical wavelengths in spectra I+V and I-V are not in the same channel, but shifted, as explained by Figure 43.

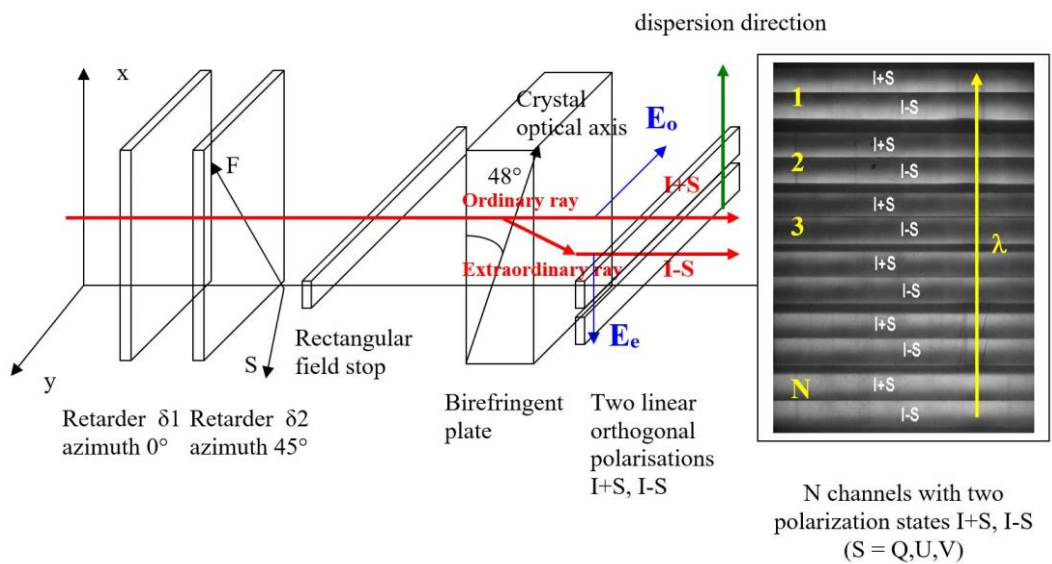

*Figure 41: dual beam polarimeter with liquid crystal retarders and calcite beam splitter-shifter of the MSDP Meudon. The rectangular field stop is the 2D entrance window of the spectrograph. Courtesy OP.*

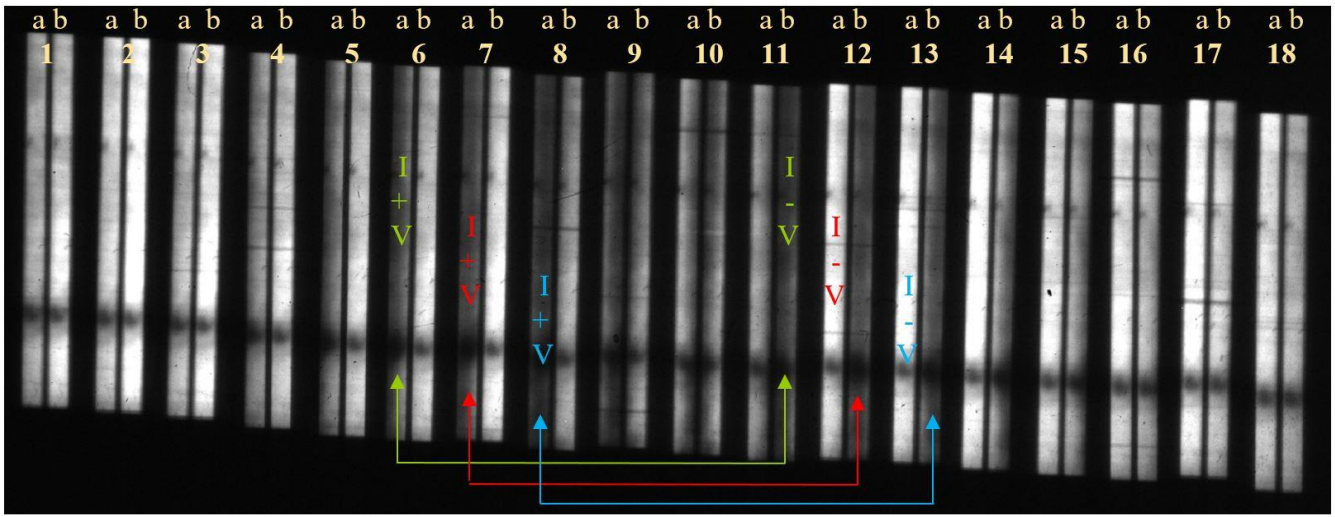

*Figure 42: dual beam circular polarimetry at the MSDP Meudon with 18 channels (or 36 subchannels I+V, I-V) in MgI 5173 Å line. Wavelength step between channels 0.029 Å. Courtesy OP.*

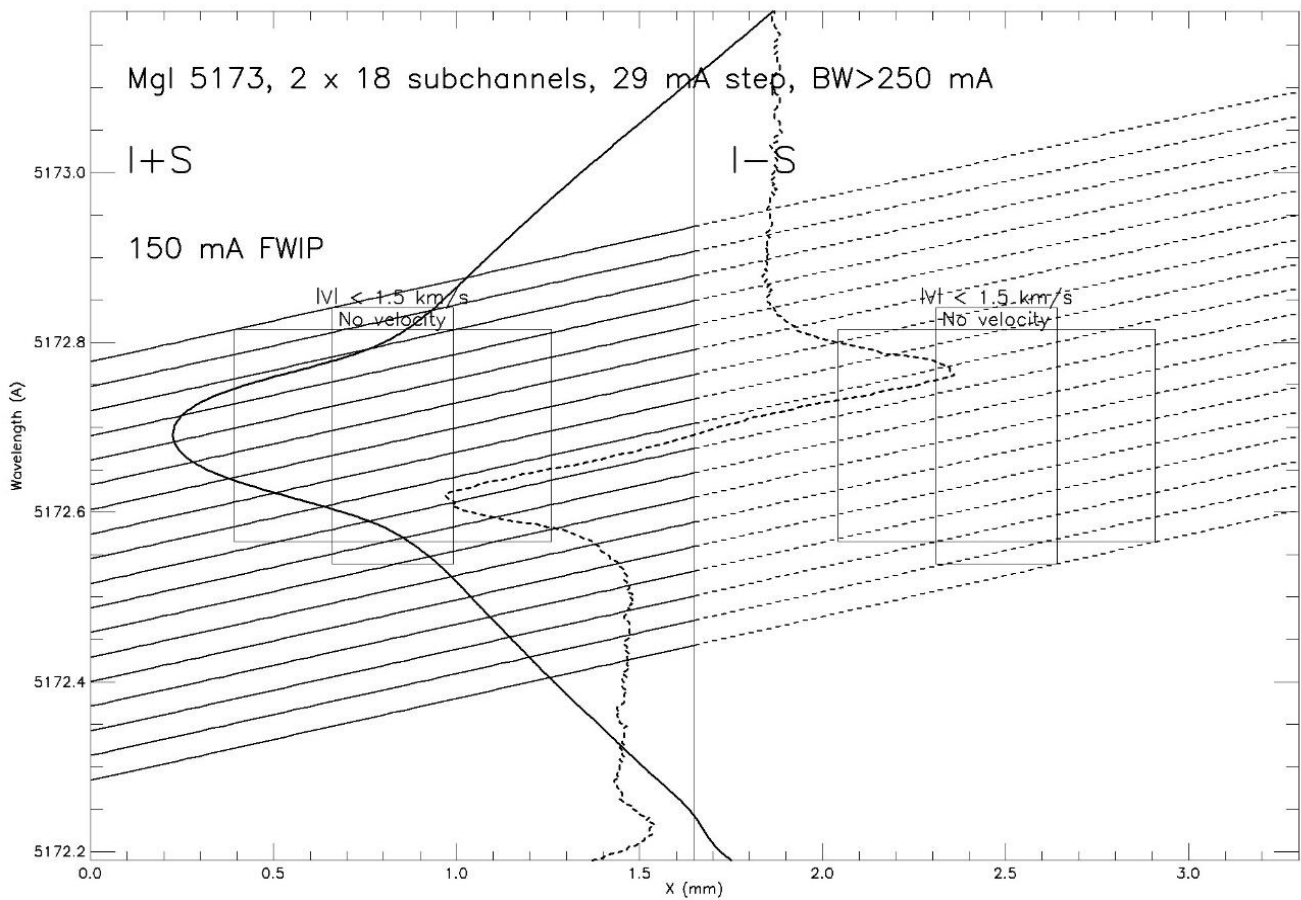

*Figure 43: wavelength transmission functions of the 18 channels of the MSDP Meudon in the case of MgI 5173 Å line. X is the abscissa (mm) along each channel; the ordinate is the wavelength (Å). The left part 0 < X < 1.65 mm corresponds to I+V; the right part 1.65 mm < X < 3.3 mm corresponds to I-V. The rectangles indicate common wavelength regions, where it is possible to restore, for each (x, y) pixel of the FOV, the spectral profiles of I+V and I-V with exactly the same wavelength sampling. The line profile is indicated (solid line) together with its first derivative (dashed line); inflexion points must appear in the two boxes to measure the LOS magnetic field B// at V/I peaks. In the presence of Dopplershifts, the V/I peaks are translated, reducing the available FOV in X direction. Courtesy OP.*

Figure 44 below is a simulation of the 18 channels/36 subchannels showing the shape of the line profiles in the case of Doppler velocities (zero and two opposite values of the Dopplershift), and in the case of longitudinal magnetic fields (I+V and I-V profiles) and transverse magnetic fields (I+Q, I-Q profiles) for two strength values (0, 1000 G). The data processing is complex and rebuilts the line profiles using the transmission functions of Figure 43 and calculates the Dopplershifts and LOS magnetic fields (Figure 45).

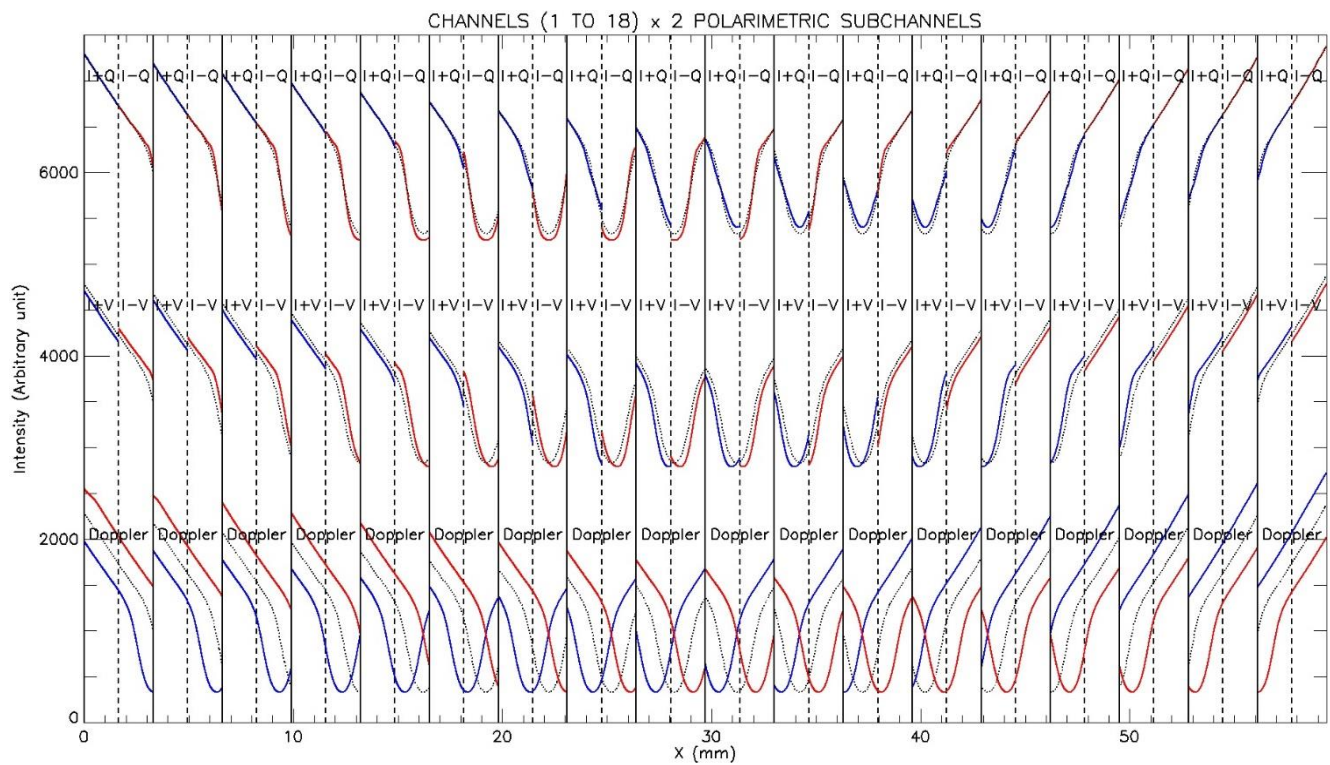

Figure 44: simulation of line profiles in the case of MgI 5173 Å line, in the presence of transverse magnetic field (top with I+Q, I-Q subchannels, 0 and 1000 G), LOS magnetic field (centre with I+V, I-V subchannels, 0 and 1000 G), and Doppler velocities (bottom). Courtesy OP.

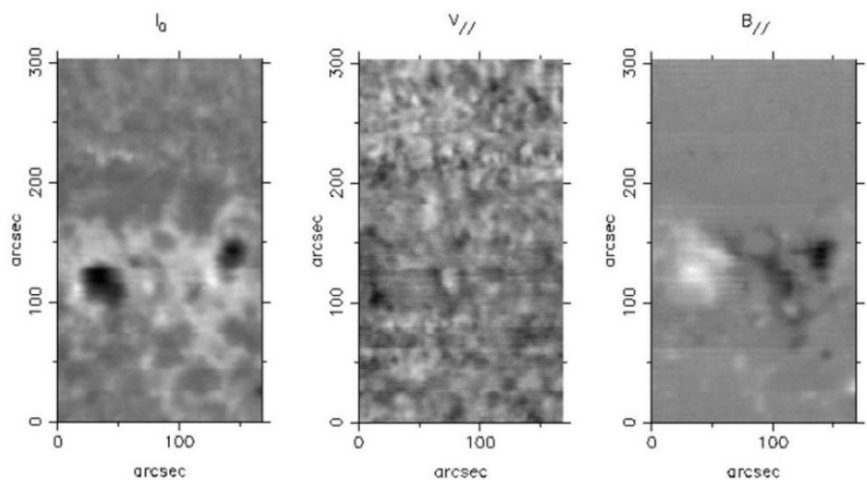

Figure 45: intensity, LOS velocity and LOS magnetic field of a sunspot in MgI 5173 Å line. Courtsey OP.

**CONCLUSION**

Many polarimetric techniques were invented and progressively developed at Meudon observatory since Lyot in order to measure solar magnetic fields with increasing temporal, spectral and spatial resolutions and improving polarimetric precision to restore the Stokes vector. The arrival in 1980 of digital detectors, such as CCD arrays, and powerful 32 bits computers (such as the mythic VAX 11/780), together with array processors (such as the Floating Point Systems AP120B) to allow fast data acquisition and handle many CCD cameras (for observations of many simultaneous spectral lines of the photosphere and chromosphere) was a revolution and allowed researchers to introduce more and more sophisticated methods of data processing. The THEMIS telescope was one of the most optimized instrument, and new polarimetric techniques such as grid masking or beam exchange with image stabilization (tip tilt or now adaptive optics) were successfully introduced, in alternative to the kHz modulation and custom cameras used by the ZIMPOL worldwide reference. The French school of polarimetry has been very productive in the second part of the twentieth century, and future telescopes in project, such as the European Solar Telescope, will greatly benefit of many advances in this domain.

## SUPPLEMENTARY DOCUMENTS

**- Movie "2cameras.mp4"**
THEMIS MTR 2-camera mode, 2 x 2' FOV, spectral lines FeI 6301 Å and FeI 6302 Å
3 successive exposures with 2 cameras
Camera 1: I+Q (exp 1), I+U (exp 2), I+V (exp 3) in sequence
Camera 2: I-Q (exp 1), I-U (exp 2), I-V (exp 3) in sequence
Beam exchange possible (not shown) with 6 successive exposures

**- Movie "1camera-beam-exchange.mp4"**
THEMIS MTR 1-camera mode, 2 x 1' FOV, spectral lines FeI around 5250 Å
6 successive exposures with a single camera (closer optical path)
I+Q, I-Q (exp 1), I+U, I-U (exp 2), I+V, I-V (exp 3), in sequence
Beam exchange:
I-Q, I+Q (exp 4), I-U, I+U (exp 5), I-V, I+V (exp 6), in sequence
Beam exchange render data independent of CCD gain table and flat field

**- Movie "grid-beam-exchange.mp4"**
THEMIS MTR 1-camera mode, grid, spectral lines FeI 6301 Å and FeI 6302 Å
3 grid steps in the FOV, 6 successive exposures (single camera)
I+Q, I-Q (exp 1), I+U, I-U (exp 2), I+V, I-V (exp 3), in sequence
Beam exchange:
I-Q, I+Q (exp 4), I-U, I+U (exp 5), I-V, I+V (exp 6), in sequence

**- Movie "MSDP-Meudon.mp4"**
18 channels x 2 = 36 polarimetric subchannels, 15" x 240" FOV
I+V, I-V simultaneous, surface scan by steps of 12"
MgI 5173 Å, step 0.029 Å between channels
CaII 8542 Å, step 0.045 Å between channels

# REFERENCES

- Arnaud, J., 1982, "observed polarization of the FeXIV 5303 coronal emission line", *Astron. Astrophys.*, 112, 350-354
- Charvin, P., 1965, "étude de la polarisation des raies interdites de la couronne solaire. Application au cas de la raie verte 5303", *Annales d'Astrophysique*, 28, 877-934
- Charvin, 1971, "Experimental study of the orientation of magnetic fields in the corona", *IAUS*, 43, 580-587
- Dollfus, A., Colson, F., Crussaire, D., Launay, F., 1985, "A monochromator for solar quantitative imagery: the instrument FPSS", *Astron. Astrophys.*, 151, 235.
- Hale, G., 1908, "On the probable existence of magnetic fields in sunspots", *ApJ*, 28, 315-343
- Hale, G., Nicholson, S., 1925, "The law of sunspot polarity", *ApJ*, 62, 270-300
- Hale, G., Nicholson, S., 1938, "Magnetic observations of sunspots", book
- Landi, E., 1992, "Magnetic field measurements, in Solar observations, techniques and instruments", *Cambridge University Press*, 73
- Leroy, J.L., 1962, "Contributions à l'étude de la polarisation de la lumière solaire", *Annales d'Astrophysique*, 25, 127-164
- Leroy, J.L., 1998, "La polarisation de la lumière et l'observation astronomique", book, Gordon and Breach science publishers
- Lyot, B., 1944, "Le filtre monochromatique polarisant et ses applications en physique solaire", *Annales d'Astrophysique*, 7, 31-79
- Malherbe, J.-M., Roudier, Th., Moity, J., Mein, P., Muller, R., 2004. "High resolution solar magnetometry with the spectrograph of the Pic du Midi turret dome". *Astron. Astrophys.*, 427, 745.
- Malherbe, J.-M., Moity, J., Arnaud, J., Roudier, Th., 2007a. "First observations of the second solar spectrum with spatial resolution at the lunette Jean Rösch". *Astron. Astrophys.*, 462, 753.
- Malherbe, J.-M., Roudier, Th., Moity, J., Mein, P., Arnaud, J., Muller, R., 2007b. "Spectropolarimetry with liquid crystals". *Mem. S. A. It.*, 78, 203.
- Mein, P., 1977. "Multichannel subtractive spectrograph and filament observations". *Solar Phys.*, 54, 45.
- Mein, P., Mein, N., Bommier, V., 2009, "Fast imaging spectroscopy with THEMIS spectrometers. Vector magnetic maps with THEMIS/MSDP", *Astron. Astrophys.*, 507, 531-539.
- Mein, P., Malherbe, J.M., Sayède, F., Rudawy, P., Phillips, K., Keenan, F., 2021. "Four decades of advances from MSDP to S4I and SLED imaging spectrometers". *Solar Phys.*, 296, 30.
- Michard, R., Mouradian, Z., Semel, M., 1961, "Champs magnétiques dans un centre d'activité solaire avant et pendant une éruption", *Annales d'Astrophysique*, 24, 54-63
- Michard, R., Rayrole, J., 1965, "Observations systématiques des champs magnétiques des centres d'activité à l'observatoire de Meudon", *IAUS*, 22, 169-172
- Mouradian, Z., Chauveau, F., Colson, F., Darré, G., Kerlirzlin, P., Olivieri, G., 1980. "The new solar spectrograph at Pic du Midi observatory". *Proceedings of the Japan France Seminar on Solar Physics,* Moryama and Hénoux editors*,* 271.
- Rayrole, J., 1985, "The new solar magnetograph for the Canary Islands Observatory THEMIS", *NASCP* 2374, 219-230
- Roudier, Th., Malherbe, J.-M., Moity, J., Rondi, S., Mein, P., Coutard, Ch., 2006. "Sub arcsec evolution of solar magnetic fields". *Astron. Astrophys.*, 455, 1091.
- Semel, M., 1980, "Un analyseur précis de polarisation optique". *Astron. Astrophys.*, 91, 369.
- Semel, M., Donati, J.F., Rees, D., 1993, Zeeman Doppler imaging of active stars", *Astron. Astrophys.*, 278, 231-237.

- Stenflo, J, 1994, "Solar magnetic fields, polarized radiation diagnostics", *Kluwer Academic Press*, book, volume 189.
- Stenflo, J., 2015, "History of solar magnetic fields since George Ellery Hale", arXiv, 1508.03312
- Stenflo, J., 2013, "solar magnetic fields as revealed by Stokes polarimetry", arXiv, 1309.5454

**THE AUTHOR**

Dr Jean-Marie Malherbe (retired in 2023) is emeritus astronomer at Paris observatory. He first worked on solar filaments and prominences using multi-wavelength observations. He used the spectrographs of the Meudon Solar Tower, the Pic du Midi Turret Dome, the German Vacuum Tower Telescope, THEMIS (Tenerife) and developed polarimeters. He proposed models and MHD 2D numerical simulations for prominence formation. More recently, he worked on the quiet Sun, using satellites such as HINODE or IRIS, and MHD simulation results. He was responsible of the Meudon spectroheliograph from 1996 to 2023 and the MeteoSpace project operating since 2024 at the Calern plateau.